\documentclass[notitlepage]{article}

\usepackage{authblk}
\usepackage{graphicx}
\usepackage{subcaption}
\usepackage{amsmath}
\usepackage{multirow}

\usepackage{xcolor}
\usepackage{url}
\usepackage{amssymb}
\usepackage{comment} 

\usepackage[T1]{fontenc}
\usepackage[utf8]{inputenc}

\usepackage{pgfplots}
\usepackage{pgfplotstable}
\usepackage{adjustbox}
\usepgfplotslibrary{groupplots}
\usepackage{tikz}
\usetikzlibrary{arrows,decorations.markings}
\usetikzlibrary{shapes.arrows}
\usetikzlibrary{patterns}
\usetikzlibrary{matrix}
\tikzstyle{internalnode} = [circle, draw, fill, minimum size=1.7mm,inner sep=0pt,outer sep=0pt]

\newtheorem{definition}{Definition}[section]

\newcommand{\myparagraph}[1]{\medskip\noindent\textbf{#1}\quad}

\newcommand\eg{\emph{e.g.}}
\newcommand\ie{\emph{i.e.}}

\newcommand{\nop}[1]{}

\def\addlegendimage{\csname pgfplots@addlegendimage\endcsname}
\selectcolormodel{cmyk}

\begin{document}
\title{Growing Graphs from Hyperedge Replacement\\Graph Grammars}
\author[1]{Salvador Aguinaga\thanks{saguinag@nd.edu}}
\author[2]{Rodrigo Palacios\thanks{rodrigopalacios@mail.fresnostate.edu}}
\author[1]{David Chiang\thanks{dchiang@nd.edu}}
\author[1]{Tim Weninger\thanks{tweninge@nd.edu}}
\affil[1]{Department of Computer Science and Engineering\\University of Notre Dame}
\affil[2]{Computer Science Department\\California State at Fresno}
\date{}                     
\setcounter{Maxaffil}{0}
\renewcommand\Affilfont{\itshape\small}
    \maketitle
\begin{abstract}
Discovering the underlying structures present in large real world graphs is a fundamental scientific problem. In this paper we show that a graph's clique tree can be used to extract a hyperedge replacement grammar. If we store an ordering from the extraction process, the extracted graph grammar is guaranteed to generate an isomorphic copy of the original graph. Or, a stochastic application of the graph grammar rules can be used to quickly create random graphs. In experiments on large real world networks, we show that random graphs, generated from extracted graph grammars, exhibit a wide range of properties that are very similar to the original graphs. In addition to graph properties like degree or eigenvector centrality, what a graph ``looks like'' ultimately depends on small details in local graph substructures that are difficult to define at a global level. We show that our generative graph model is able to preserve these local substructures when generating new graphs and performs well on new and difficult tests of model robustness.
\end{abstract}

\tikzset{textnode/.style={inner sep=0pt,outer sep=0,execute at begin node={\strut}}}
\tikzstyle{opencircle} = [circle,minimum width=10, draw, fill=black!5, inner sep=1.5]
\tikzstyle{faintopencircle} = [circle,minimum width=10, draw=black!40, fill=black!05, inner sep=1.5, text=black!40]
\tikzstyle{itxset} = [rounded corners=3pt, draw, minimum height=14pt, inner sep=0]
\tikzstyle{itxsetfocus} = [itxset, ultra thick]
\tikzstyle{internalnode} = [textnode,circle, draw, fill, text=white]
\tikzstyle{externalnode} = [textnode, circle, draw]
\tikzstyle{nonterminal} = [textnode,draw,inner xsep=1.5]
\tikzstyle{graphletnode} = [circle, draw, fill, minimum size=1.0mm,inner sep=0pt,outer sep=0pt]
\tikzstyle{splitnode} = [circle, draw, fill, minimum size=1.2mm,inner sep=0pt,outer sep=0pt]
\tikzset{ghost/.style={color=gray!50}}
\tikzstyle{faintnonterminal} = [draw=black!40, inner sep=1.5, text=black!40]
\tikzstyle{faintedge} = [very thin, color=black!50]

\section{Introduction}

Teasing out signatures of interactions 
buried in overwhelming volumes of information is one of the most basic challenges in scientific research. Understanding how information is organized can help us discover its fundamental underlying properties. Researchers do this when they investigate the relationships between diseases, cell functions, chemicals, or particles, and we all learn new concepts and solve problems by understanding the relationships between the various entities present in our everyday lives. These entities can be represented as networks, or graphs, in which local behaviors can be easily understood, but whose global view is highly complex. 

These networks exhibit a long and varied list of global properties, including heavy-tailed degree distributions~\cite{strogatz2001exploring} and interesting growth characteristics~\cite{leskovec2005graphs,LeskovecKleinbergFaloutsos:2007}, among others. Recent work has found that these global properties are merely products of a graph's local properties, in particular, graphlet distributions~\cite{Ugander2013}. These small, local substructures often reveal the degree distributions, diameter and other global properties of a graph~\cite{przulj2007biological, Ugander2013}, and have been shown to be a more complete way to measure the similarity between two or more graphs~\cite{yaverouglu2015proper}. Our overall goal, and the goal of structural inference algorithms in general, is to learn the local structures that, in aggregate, help describe the observed interactions and generalize to explain further phenomena. 

For example, physicists and chemists have found that many chemical interactions are the result of underlying structural properties of the individual elements. Similarly, biologists have agreed that simple tree structures are useful when organizing the evolutionary history of life, and sociologists find that clique-formation, \eg, triadic closure, underlies community development~\cite{Canestro2007Nature, Sadava:2014, backstrom2006group}. In other instances, the structural organization of the entities may resemble a ring, a clique, a star, or any number of complex configurations.

In this work, we describe a general framework that can discover, from any large network, simple structural forms in order to make predictions about the topological properties of a network. In addition, this framework is able to extract mechanisms of network generation from small samples of the graph in order to generate networks that satisfy these properties. 
Our major insight is that a network's {\em clique tree} encodes simple information about the structure of the network. We use the closely-related formalism of \emph{hyperedge replacement grammars} (HRGs) as a way to describe the organization of real world networks. 

Unlike previous models that manually define the space of possible structures~\cite{Kemp05082008} or define the grammar by extracting frequent subgraphs~\cite{kukluk2006inference,kukluk2008inference}, our framework can automatically discover the necessary forms and use them to recreate the original graph {\em exactly} as well as infer generalizations of the original network. Our approach can handle any type of graph and does not make any assumption about the topology of the data. 

After reviewing some of the theoretical foundations of clique trees and HRGs, we  show how to extract an HRG from a graph and use it to reconstruct the original graph. We then show how to use the extracted grammar to stochastically generate generalizations of the original graph. Finally, we present experimental results that compare the stochastically generated graphs with the original graphs. We show that these generated graphs exhibit a wide range of properties that are very similar to the properties of the original graphs, and significantly outperform existing graph models at generating subgraph distributions similar to those found in the original~graph.

\section{Preliminaries}

Before we describe our method, some background definitions are needed. We begin with an arbitrary input {\em hypergraph} $H=(V,E)$, where a {\em hyperedge} $e\in E$ can connect multiple vertices $e_i=\{v_1,v_2,\ldots,v_k\}$. Common {\em graphs} (\eg, social networks, Web graphs, information networks) are a particular case of hypergraphs where each edge connects exactly two vertices. For convenience, all of the graphs in this paper will be {\em simple}, {\em connected} and {\em undirected}, although these restrictions are not vital. In the remainder of this section we refer mainly to previous developments in clique trees and their relationship to hyperedge replacement grammars in order to support the claims made in sections 3 and 4.

\subsection{Clique Trees}

All graphs can be decomposed (though not uniquely) into a {\em clique tree}, also known as a tree decomposition, junction tree, join tree, intersection tree, or cluster graph.
Within the data mining community, clique trees are best known for their role in exact inference in probabilistic graphical models, so we introduce the preliminary work from a graphical modelling perspective; for an expanded introduction, we refer the reader to Chapters 9 and 10 of Koller and Friedman's textbook~\cite{Koller:2009:PGM:1795555}.

\begin{definition}
\label{defn:cliquetree}
A {\em clique tree} of a graph $H=(V,E)$ is a 
tree $T$, each of whose nodes $\eta$ is labeled with a $V_{\eta}\subseteq V$ and $E_{\eta}\subseteq E$, such that the following properties hold:
\begin{enumerate}
\item Vertex Cover: For each $v\in V$, there is a vertex $\eta\in T$ such that $v\in V_\eta$. 
\item Edge Cover: For each hyperedge $e_i=\{v_1,\ldots,v_k\} \in E$ there is exactly one node $\eta\in T$ such that $e\in E_\eta$. Moreover, $v_1,\ldots,v_k\in V_\eta$.
\item Running Intersection: For each $v\in V$, the set $\{\eta\in T \mid v\in V_\eta\}$ is connected.
\end{enumerate}
\end{definition}

\begin{definition}
The \emph{width} of a clique tree is $\max(|V_{\eta}-1|)$, and the \emph{treewidth} of a graph $H$ is the minimal width of any clique tree of $H$.
\end{definition}
Unfortunately, finding the optimal elimination ordering and corresponding minimal-width clique tree is NP-complete \cite{Arnborg1987}. Fortunately, many reasonable approximations exist for general graphs: in this paper we employ the commonly used maximum cardinality search (MCS) heuristic introduced by Tarjan and Yannikakis~\cite{TarjanYannakakis1985} in order to compute a clique tree with a reasonably-low, but not necessarily minimal, width.

Simply put, a clique tree of any graph (or any hypergraph) is a tree, each of whose nodes is labeled with nodes and edges from the original graph, such that {\em vertex cover}, {\em edge cover} and the {\em running intersection} properties hold, and the ``width'' of the clique tree measures how tree-like the graph is. The reason for the interest in finding the clique tree of a graph is because many computationally difficult problems can be solved efficiently when the data is constrained to be a tree.

\begin{figure}[t]
\centering
\includegraphics{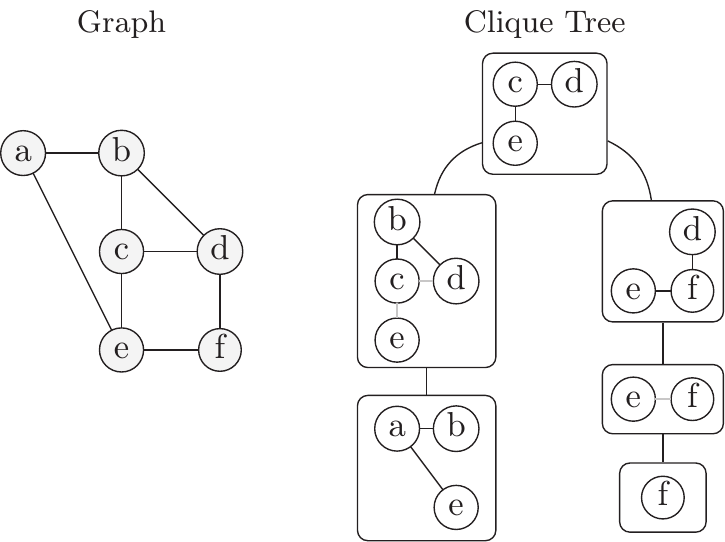}
\caption{A graph and one possible minimal-width clique tree for it. Ghosted edges are not part of $E_\eta$; they are shown only for explanatory purposes.
}
\label{fig:expdtree}
\end{figure}

Figure~\ref{fig:expdtree} shows a graph and a minimal-width clique tree of the same graph (showing $V_\eta$ for each node $\eta$). Nodes are labeled with lowercase Latin letters. We will refer back to this graph and clique tree as a running example throughout this paper.

\subsection{Hyperedge Replacement Grammar}
The key insight for this task is that a network's clique tree encodes robust and precise information about the network. An HRG, which is extracted from the clique tree, contains graphical rewriting rules that can match and replace graph fragments similar to how a Context Free Grammar (CFG) rewrites characters in a string. These graph fragments represent a succinct, yet complete description of the building blocks of the network, and the rewriting rules of the HRG represent the instructions on how the graph is pieced together. For a thorough examination of HRGs, we refer the reader to the survey by Drewes {\em et al.}~\cite{Drewes2002}.

\begin{definition}
\label{defn:tuple}
A {\em hyperedge replacement grammar} is a tuple $G=\langle N,T,S,\mathcal{P}\rangle$, where
\begin{enumerate}
\item $N$ is a finite set of nonterminal symbols. Each nonterminal $A$ has a nonnegative integer \emph{rank}, which we write $|A|$.
\item $T$ is a finite set of terminal symbols.
\item $S\in N$ is a distinguished starting nonterminal, and $|S|=0$.
\item $\mathcal{P}$ is a finite set of production rules $A\rightarrow R$, where 
\begin{itemize}
\item $A$, the left hand side (LHS), is a nonterminal symbol.
\item $R$, the right hand side (RHS), is a hypergraph whose edges are labeled by symbols from $T\cup N$. If an edge $e$ is labeled by a nonterminal $B$, we must have $|e| = |B|$.
\item Exactly $|A|$ vertices of $R$ are designated \emph{external vertices}. The other vertices in $R$ are called \emph{internal} vertices.
\end{itemize}
\end{enumerate}
\end{definition}
When drawing HRG rules, we draw the LHS $A$ as a hyperedge labeled $A$ with arity $|A|$. We draw the RHS as a hypergraph, with the external vertices drawn as solid black circles and the internal vertices as open white circles.

If an HRG rule has no nonterminal symbols in its RHS, we call it a \emph{terminal rule}.

\begin{definition}
\label{defn:hrg}
Let $G$ be an HRG and $P=(A \rightarrow R)$ be a production rule of $G$. We define the relation $H'\Rightarrow H^{*}$ ($H^{*}$ is derived in one step from $H'$) as follows. 
$H'$ must have a hyperedge $e$ labeled $A$; let $v_1, \ldots, v_k$ be the vertices it connects. Let $u_1, \ldots, u_k$ be the external vertices of $R$. Then $H^*$ is the graph formed by removing $e$ from $H'$, making an isomorphic copy of $R$, and identifying $v_i$ with the copies of $u_i$ for each $i=1,\ldots,k$.

Let $\Rightarrow^\ast$ be the reflexive, transitive closure of $\Rightarrow$. Then we say that $G$ generates a graph $H$ if there is a production $S \rightarrow R$ and $R \Rightarrow^\ast H$ and $H$ has no edges labeled with nonterminal symbols.
\end{definition}

In other words, a derivation starts with the symbol $S$, and we repeatedly choose a  nonterminal $A$ and rewrite it using a production $A \rightarrow R$. The replacement hypergraph fragments $R$ can itself have other nonterminal hyperedges, so this process is repeated until there are no more nonterminal hyperedges.

These definitions will be clearly illustrated in the following sections.

\subsection{The Missing Link}
Clique trees and hyperedge replacement graph grammars have been studied for some time in discrete mathematics and graph theory literature. HRGs are conventionally used to generate graphs with very specific structures, \eg, rings, trees, stars. A drawback of many current applications of HRGs is that their production rules must be hand drawn to generate some specific graph or class of graphs. Very recently, Kemp and Tenenbaum developed an inference algorithm that learned probabilities from real world graphs, but still relied on a handful of rather basic hand-drawn production rules (of a related formalism called vertex replacement grammar) to which the learned probabilities were assigned~\cite{Kemp05082008}.

The main contribution of this paper is to combine prior theoretical work on clique trees, tree decomposition and treewidth to automatically learn an HRG for real world graphs. Existing graph generators, like exponential random graphs, small world graphs, Kronecker graphs, and so on, learn parameters from some input graph to generate new graphs stochastically. Unlike these previous approaches, our model has the ability to reproduce the exact same graph topology where the new graph is guaranteed to be isomorphic to the original graph. Our model is also able to stochastically generate different-sized graphs that share similar properties to the original graph.

\section{Learning Graph Grammars}

The first step in learning an HRG from a graph is to compute a clique tree from the original graph. Then, this clique tree induces an HRG in a natural way, which we demonstrate in this section.

\subsection{Clique Trees and HRGs}

Let $\eta$ be an interior node of the clique tree $T$, let $\eta'$ be its parent, and let $\eta_1, \ldots, \eta_m$ be its children. Node $\eta$ corresponds to an HRG production rule $A \rightarrow R$ as follows. First, $|A| = |V_{\eta'} \cap V_{\eta}|$. Then, $R$ is formed by:
\begin{itemize}
\item Adding an isomorphic copy of the vertices in $V_\eta$ and the edges in $E_\eta$
\item Marking the (copies of) vertices in $V_{\eta'} \cap V_{\eta}$ as external vertices
\item Adding, for each $\eta_i$, a nonterminal hyperedge connecting the (copies of) vertices in $V_\eta \cap V_{\eta_i}$.
\end{itemize}

\begin{figure}[h]
\centering
\includegraphics{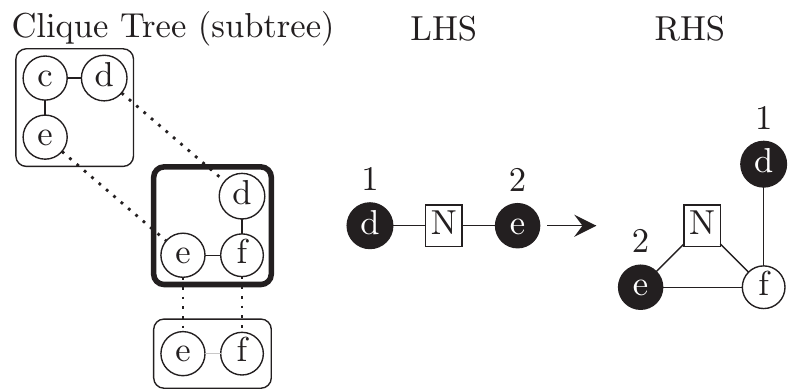}
\caption{Example of hyperedge replacement grammar rule creation from an interior vertex of the clique tree. Note that lowercase letters inside vertices are for explanatory purposes only; only the numeric labels outside external vertices are actually part of the rule.}
\label{fig:creation}
\end{figure}

Figure~\ref{fig:creation} shows an example of the creation of an HRG rule. In this example, we focus on the middle clique-tree node $V_\eta = \{\text{d},\text{e},\text{f}\}$, outlined in bold. 

We choose nonterminal symbol N for the LHS, which must have rank~2 because $\eta$ has 2 vertices in common with its parent.
The RHS is a graph whose vertices are (copies of) $V_{\eta} = \{\text{d}, \text{e}, \text{f}\}$. Vertices d and e are marked external (and numbered 1 and 2, arbitrarily) because they also appear in the parent node. The terminal edges are $E_{\eta} = \{(\text{d},\text{f}), (\text{e},\text{f})\}$. There is only one child of $\eta$, and the nodes they have in common are e and~f, so there is one nonterminal hyperedge connecting e and~f.

Next we deal with the special cases of the root and leaves.

\begin{figure}[h]
\centering
\includegraphics{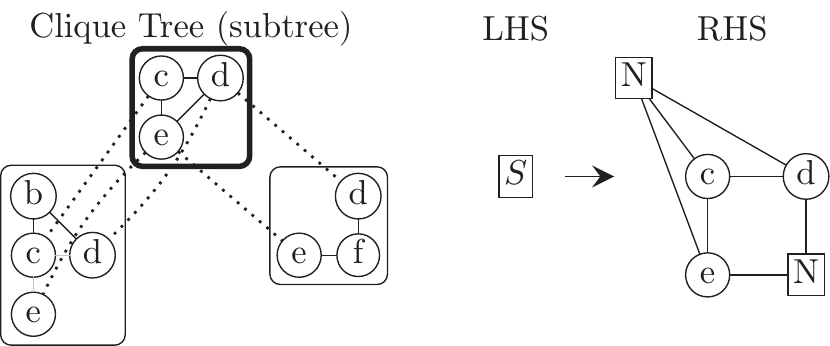}
\caption{Example of hyperedge replacement grammar rule creation from the root node of the clique tree.}
\label{fig:creation_root}
\end{figure}

\myparagraph{Root Node.} If $\eta$ is the root node, then it does not have any parent cliques, but may still have one or more children. Because $\eta$ has no parent, the corresponding rule has a LHS with rank 0 and a RHS with no external vertices. In this case, we use the start nonterminal $S$ as the LHS, as shown in Figure~\ref{fig:creation_root}.

The RHS is computed in the same way as the interior node case. For the example in Fig.~\ref{fig:creation_root}, the RHS has vertices that are copies of c, d, and e. In addition, the RHS has two terminal hyperedges, $E_\eta = \{(\text{c},\text{d}),(\text{c},\text{e})\}$. The root node has two children, so there are two nonterminal hyperedges on the RHS. The right child has two vertices in common with $\eta$, namely, d and e; so the corresponding vertices in the RHS are attached by a 2-ary nonterminal hyperedge. The left child has three vertices in common with $\eta$, namely, c, d, and e, so the corresponding vertices in the RHS are attached by a 3-ary nonterminal hyperedge.

\begin{figure}[h]
\centering
\includegraphics{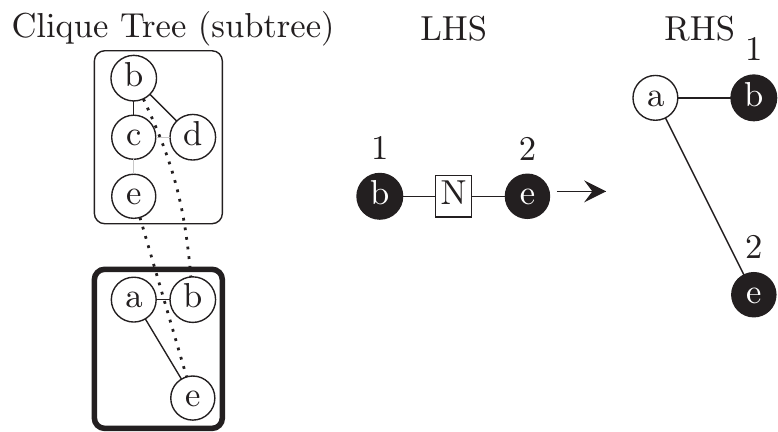}
\caption{Example of hyperedge replacement grammar rule creation from a leaf vertex of the clique tree.}
\label{fig:creation_leaf}
\end{figure}

\myparagraph{Leaf Node.} If $\eta$ is a leaf node, then the LHS is calculated the same as in the interior node case. Again we return to the running example in Fig.~\ref{fig:creation_leaf} (on the next page). Here, we focus on the leaf node $\{\text{a},\text{b},\text{e}\}$, outlined in bold. The LHS has rank 2, because $\eta$ has two vertices in common with its parent.

The RHS is computed in the same way as the interior node case, except no new nonterminal hyperedges are added to the RHS. The vertices of the RHS are (copies of) the nodes in $\eta$, namely, a, b, and e. Vertices b and e are external because they also appear in the parent clique. This RHS has two terminal hyperedges, $E_\eta = \{(\text{a},\text{b}), (\text{a},\text{e})\}$. Because the leaf clique has no children, it cannot produce any nonterminal hyperedges on the RHS; therefore this rule is a terminal rule.

\subsection{Top-Down HRG Rule Induction} 
We induce production rules from the clique tree by applying the above extraction method top down. Because trees are acyclic, the traversal order does not matter, yet there are some interesting observations we can make about traversals of moderately sized graphs. First, exactly one HRG rule will have the special starting nonterminal $S$ on its LHS; no mention of $S$ will ever appear in any RHS. Similarly, the number of terminal rules is equal to the number of leaf nodes in the clique tree.

Larger graphs will typically produce larger clique trees, especially sparse graphs because they are more likely to have a larger number of small maximal cliques. These larger clique trees will produce a large number of HRG rules, one for each clique in the clique tree. Although it is possible to keep track of each rule and its traversal order, we find, and will later show in the experiments section, that the same rules are often repeated many times.

\renewcommand{\boxed}[1]{\text{\fboxsep=.1em\fbox{$\displaystyle#1$}}}

\begin{figure}[t]
\centering
\resizebox{.50\hsize}{!}{%
\includegraphics{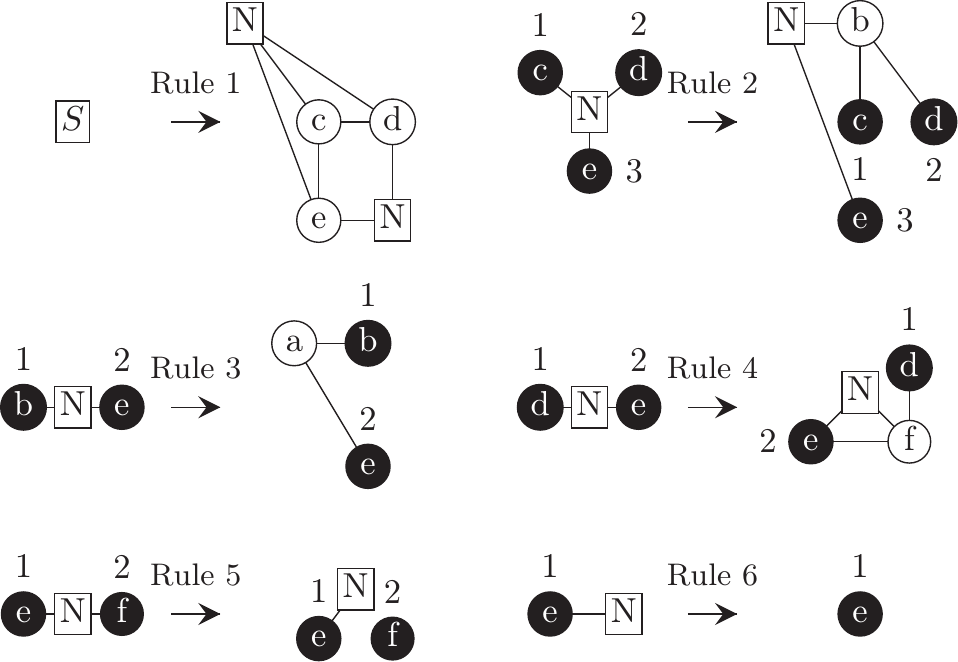}
}
\caption{Complete set of production rules extracted from the example clique tree. Note that lowercase letters inside vertices are for explanatory purposes only; only the numeric labels outside external vertices are actually part of the rule.}
\label{fig:production_rules}
\end{figure}

Figure~\ref{fig:production_rules} shows the 6 rules that are induced from the clique trees illustrated in Fig.~\ref{fig:expdtree} and used in the running example throughout this section.

\subsection{Complexity Analysis} 
The HRG rule induction steps described in this section can be broken into two steps: 
(i) creating a clique tree and (ii) the HRG rule extraction process. 

Unfortunately, finding a clique tree with minimal width \ie, the treewidth $tw$, is NP-Complete. Let $n$ and $m$ be the number of vertices and edges respectively in $H$. Tarjan and Yannikakis' Maximum Cardinality Search (MCS) algorithm finds a usable clique tree~\cite{TarjanYannakakis1985} in linear time $O(n+m)$, but is not guaranteed to be minimal.

The running time of the HRG rule extraction process is determined exclusively by the size of the clique tree as well as the number of vertices in each clique tree node. From Defn.~\ref{defn:cliquetree} we have that the number of nodes in the clique tree is $m$. When minimal, the number of vertices in an the largest clique tree node $\max(|\eta_i|)$ (minus 1) is defined as the treewidth $tw$, however, clique trees generated by MCS have $\max(|\eta_i|)$ bounded by the maximum degree of $H$, denoted as $\Delta$~\cite{geman2002dynamic}. Therefore, given an elimination ordering from MCS, the computational complexity of the extraction process is in $O(m\cdot \Delta)$.

\section{Graph Generation}

In this section we show how to use the HRG extracted from the original graph $H$ (as described in the previous section) to generate a new graph $H^*$. Ideally, $H^*$ will be similar to, or have features that are similar to the original graph $H$. We present two generation algorithms. The first generation algorithm is {\em exact generation}, which, as the name implies, creates an isomorphic copy of the original graph $H^*\equiv H$. The second generation algorithm is a fast {\em stochastic generation} technique that generates random graphs with similar characteristics to the original graph. 

Each generation algorithm starts with $H^\prime$ containing only the starting nonterminal $S$.

\subsection{Exact Generation}
Exact generation operates by reversing the HRG extraction process. In order to do this, we must store the HRG rules $\mathcal{P}$ as well as the clique tree $T$ (or at least the order that the rules were created). 
The first HRG rule considered is always the rule with the nonterminal labelled $S$ as the LHS. This is because the clique tree traversal starts at the root, and because the root is the only case that results in $S$ on the LHS. 

The previous section defined an HRG $G$ that is constructed from a clique tree $T$ of some given hypergraph $H$, and Defn.~\ref{defn:hrg} defines the application of a production rule $(A\rightarrow R)$ that transforms some hypergraph $H^{\prime}$ into a new hypergraph $H^*$. By applying the rules created from the clique tree in order, we will create an $H^*$ that is  isomorphic to the original hypergraph $H$. 

In the remainder of this section, we provide a more intuitive look at the exact generation property of the HRG by recreating the graph decomposed in the running example. 

\begin{figure}[h]
\centering
\includegraphics{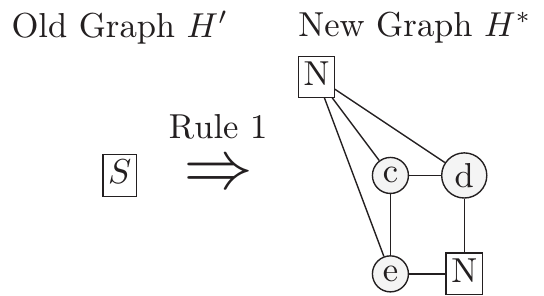}
\caption{Application of Rule 1 to replace the starting nonterminal $S$ with the RHS to create a new graph $H^*$.}
\label{fig:rule1}
\end{figure}

Using the running example from the previous section, the application of Rule 1 illustrated in Fig.~\ref{fig:rule1} shows how we transform the starting nonterminal into a new hypergraph, $H^*$. This hypergraph now has two nonterminal hyperedges corresponding to the two children that the root clique had in Fig.~\ref{fig:expdtree}. The next step is to replace $H^\prime$ with $H^*$ and then pick a nonterminal corresponding to the leftmost unvisited node of the clique tree.

\begin{figure}[h]
\centering
\includegraphics{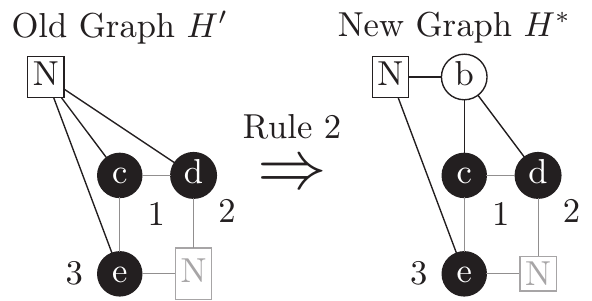}
\caption{Application of Rule 2 to replace a size-3 nonterminal in $H^\prime$ with the RHS to create a new graph $H^*$.}
\label{fig:rule2}
\end{figure}

We proceed down the left hand side of the clique tree, applying Rule 2 to $H^\prime$ as shown in Fig.~\ref{fig:rule2}. The LHS of Rule 2 matches the 3-ary hyperedge and replaces it with the RHS, which introduces a new internal vertex, two new terminal edges and a new nonterminal hyperedge. Again we set $H^\prime$ to be $H^*$ and continue to the leftmost leaf in the example clique tree.

\begin{figure}[h]
\centering
\includegraphics{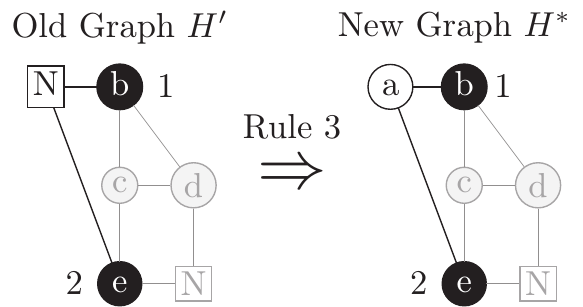}
\caption{Application of Rule 3 to replace a size-2 nonterminal in $H^\prime$ with the RHS to create a new graph $H^*$.}
\label{fig:rule3}
\end{figure}

The leftmost leaf in Fig.~\ref{fig:expdtree} corresponds to the application of Rule~3; it is the next to be applied to the new nonterminal in $H^*$ and replaced by the RHS as illustrated in Figure~\ref{fig:rule3}. The LHS of Rule 3 matches the 2-ary hyperedge shown and replaces it with the RHS, which creates a new internal vertex along with two terminal edges. Because Rule 3 comes from a leaf node, it is a terminal rule and therefore does not add any nonterminal hyperedges. This concludes the left subtree traversal from Fig.~\ref{fig:expdtree}. 

\begin{figure}[h]
\centering
\includegraphics{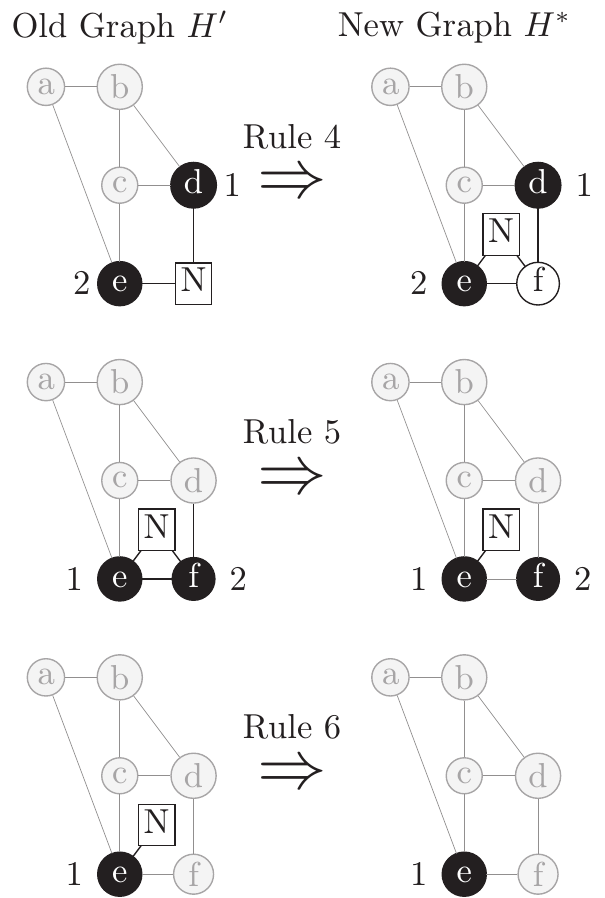}
\caption{Application of Rules 4, 5 and 6 to create an $H^*$ that is isomorphic to the original graph $H$.}
\label{fig:rule456}
\end{figure}

Continuing the example, the right subtree in the clique tree illustrated in Fig.~\ref{fig:expdtree} has three further applications of the rules in $\mathcal{P}$. As illustrated in Fig.~\ref{fig:rule456}, Rule 4 adds the final vertex, two terminal edges and one nonterminal hyperedge to $H^*$. Rule 5 and Rule 6 do not create any more terminal edges or internal vertices in $H^*$, but are still processed because of the way the clique tree is constructed.

After all 6 rules are applied in order, we are guaranteed that $H$ and $H^*$ are isomorphic.

\subsection{Stochastic Generation}
There are many cases in which we prefer to create very large graphs in an efficient manner that still exhibit the local and global properties of some given example graph {\em without storing the large clique tree} as required in exact graph generation.  Here we describe a simple stochastic hypergraph generator that applies rules from the extracted HRG in order to efficiently create graphs of arbitrary size.

In larger HRGs we usually find many $A\rightarrow R$ production rules that are identical. We can merge these duplicates by matching rule-signatures in a dictionary, and keep a count of the number of times that each distinct rule has been seen. For example, if there were some additional Rule 7 in Fig.~\ref{fig:production_rules} that was identical to, say, Rule 3, then we would simply note that we saw Rule 3 two times.

To generate random graphs from a probabilistic HRG, we start with the special starting nonterminal $H^\prime = S$. From this point, $H^*$ can be generated as follows: (1) Pick any nonterminal $A$ in $H^\prime$; (2) Find the set of rules $(A \rightarrow R)$ associated with LHS $A$; (3) Randomly choose one of these rules with probability proportional to its count; (4) replace $A$ in $H^\prime$ with $R$ to create $H^*$; (5) Replace $H^\prime$ with $H^*$ and repeat until there are no more nonterminal edges. 

However, we find that although the sampled graphs have the same mean size as the original graph, the variance is much too high to be useful. So we want to sample only graphs whose size is the same as the original graph's, or some other user-specified size. Naively, we can do this using rejection sampling: sample a graph, and if the size is not right, reject the sample and try again. However, this would be quite slow. Our implementation uses a dynamic programming approach to do this exactly while using quadratic time and linear space, or approximately while using linear time and space. We omit the details of this algorithm here, but the source code is available online at \url{https://github.com/nddsg/HRG/}.

\section{Experiments}

HRGs contain rules that succinctly represent the global and local structure of the original graph. In this section, we compare our approach against some of the state-of-the-art graph generators. We consider the properties that underlie a number of real-world networks and compare the distribution of graphs generated using generators for Kronecker Graphs, the Exponential Random Graph, Chung-Lu Graphs, and the graphs produced by the stochastic hyperedge replacement graph grammar. 

In a manner similar to HRGs, the Kronecker and Exponential Random Graph Models learn parameters that can be used to approximately recreate the original graph $H$ or a graph of some other size such that the stochastically generated graph holds many of the same properties as the original graph. The Chung-Lu graph model relies on node degree sequences to yield graphs that maintain this distribution. Except in the case of exact HRG generation described above, the stochastically generated graphs are likely not isomorphic to the original graph. We can, however, still judge how closely the stochastically generated graph resembles the original graph by comparing several of their properties. 

\subsection{Real World Datasets}
In order to get a holistic and varied view of the strengths and weaknesses of HRGs in comparison to the other leading graph generation models, we consider real-world networks that exhibit properties that are both common to many networks across different fields, but also have certain distinctive properties. 

\begin{table}[h]
\vspace{-.5cm}
\centering
\caption{Real networks}
\begin{tabular}{r|cc}
  \textbf{Dataset Name} & \textbf{Nodes} & \textbf{Edges} \\\hline
  Enron Emails & 36,692 & 183,831 \\
  ArXiv GR-QC & 5,242 & 14,496\\
  Internet Routers & 6,474 & 13,895\\
  DBLP & 317,080 & 1,049,866\\
\end{tabular}
\label{tab:realnets}
\vspace{-.2cm}
\end{table}

The four real world networks considered in this paper are described in Table~\ref{tab:realnets}. The networks vary in their number of vertices and edges as indicated, but also vary in clustering coefficient, diameter, degree distribution and many other graph properties. Specifically, the Enron graph is the email correspondence graph of the now defunct Enron corporation; the ArXiv GR-QC graph is the co-authorship graph extracted from the General Relativity and Quantum Cosmology section of ArXiv; the Internet router graph is created from traffic flows through Internet peers; and, finally, DBLP is the co-authorship graph from the DBLP dataset. Datasets were downloaded from the SNAP and KONECT dataset repositories.

\subsection{Methodology}
\label{sec:methodology}

We compare several different graph properties from the 4 classes of graph generators (HRG, Kronecker, Chung-Lu and exponential random graph (ERGM) models) to the original graph $H$. Other models, such as the Erd\H{o}s-R\'{e}nyi random graph model, the Watts-Strogatz small world model, the Barab\'{a}si-Albert generator, etc. are not compared here due to limited space and because Kronecker, Chung-Lu and ERGM have been shown to outperform these earlier models when matching network properties in empirical networks.

Kronecker graphs operate by learning an initiator matrix and then performing a recursive multiplication of that initiator matrix in order to create an adjacency matrix of the approximate graph. In our case, we use KronFit~\cite{Leskovec2010kronecker} with default parameters to learn a $2\times 2$ initiator matrix and then use the recursive Kronecker product to generate the graph. Unfortunately, the Kronecker product only creates graphs where the number of nodes is a power of 2, \ie, $2^x$, where we chose $x=15$, $x=12$, $x=13$, and $x=18$ for Enron, ArXiv, Routers and DBLP graphs respectively to match the number of nodes as closely as possible.

The Chung-Lu Graph Model (CL) takes, as input, a degree distribution and generates a new graph of the similar degree distribution and size~\cite{ChungLu2002connected}.

Exponential Random Graph Models (ERGMs) are a class of probabilistic models used to directly describe several structural features of a graph~\cite{Robins2007}. We used default parameters in R's ERGM package~\cite{Hunter2008} to generate graph models for comparison. In addition to the problem of model degeneracy, ERGMs do not scale well to large graphs. As a result, DBLP and Enron could not be modelled due to their size, and the ArXiv graph always resulted in a degenerate model. Therefore ERGM results are omitted from this section.

The main strength of HRG is to learn the patterns and rules that generate a large graph from only a few small subgraph-samples of the original graph. So, in all experiments, we make $k$ random samples of size $s$ node-induced subgraphs by a breadth first traversal starting from a random node in the graph~\cite{Leskovec2006sampling}. By default we set $k=4$ and $s=500$ empirically. We then compute tree decompositions from the $k$ samples, learn HRGs $G_1, G_2,\ldots, G_{k}$, and combine them to create a single grammar $G = \bigcup_i G_i$. For evaluation purposes, we generate 20 approximate graphs for the HRG, Chung-Lu, and Kronecker models and plot the mean values in the results section. We did compute the confidence internals for each of the models, but omitted them from the graphs for clarity. In general, the confidence intervals were very small for HRG, Kronecker and CL (indicating good consistency), but very big in the few ERGM graphs that we were able to generate because of the model degeneracy problem we encountered.

\subsection{Graph Generation Results}

Here we compare and contrast the results of approximate graphs generated from HRG, Kronecker product, and Chung-Lu. Before the results are presented, we briefly introduce the graph properties that we use to compare the similarity between the real networks and their approximate counterparts. Although many properties have been discovered and detailed in related literature, we focus on three of the principal properties from which most others can be derived.

\begin{figure}[t]
\centering
\begin{minipage}[b]{0.23\textwidth}
    \resizebox{\textwidth}{!}{
        \includegraphics{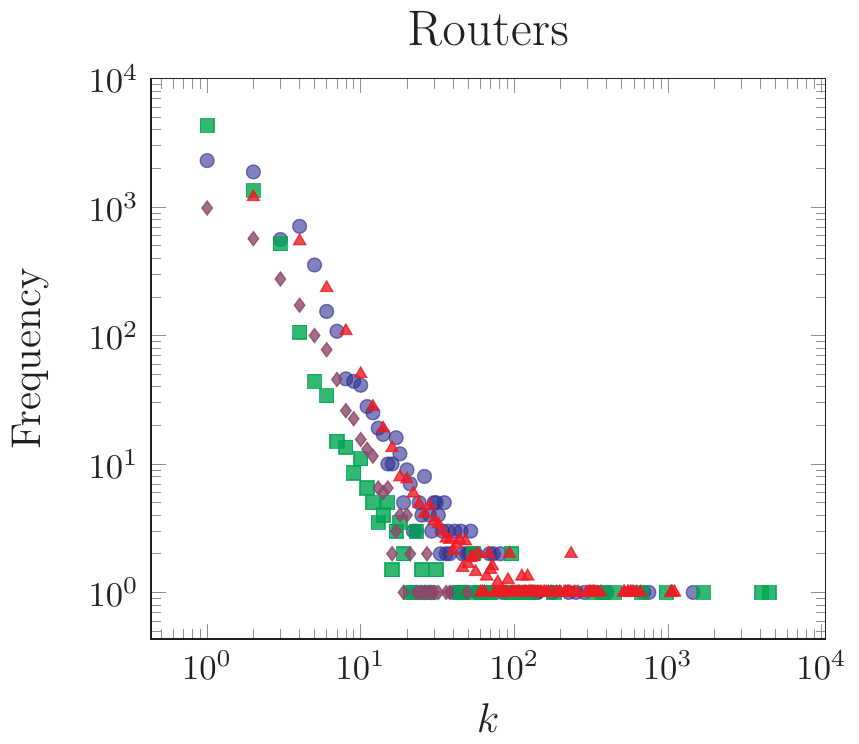}
    }
\end{minipage}%
\begin{minipage}[b]{0.222\textwidth}
    \resizebox{\textwidth}{!}{
        \includegraphics{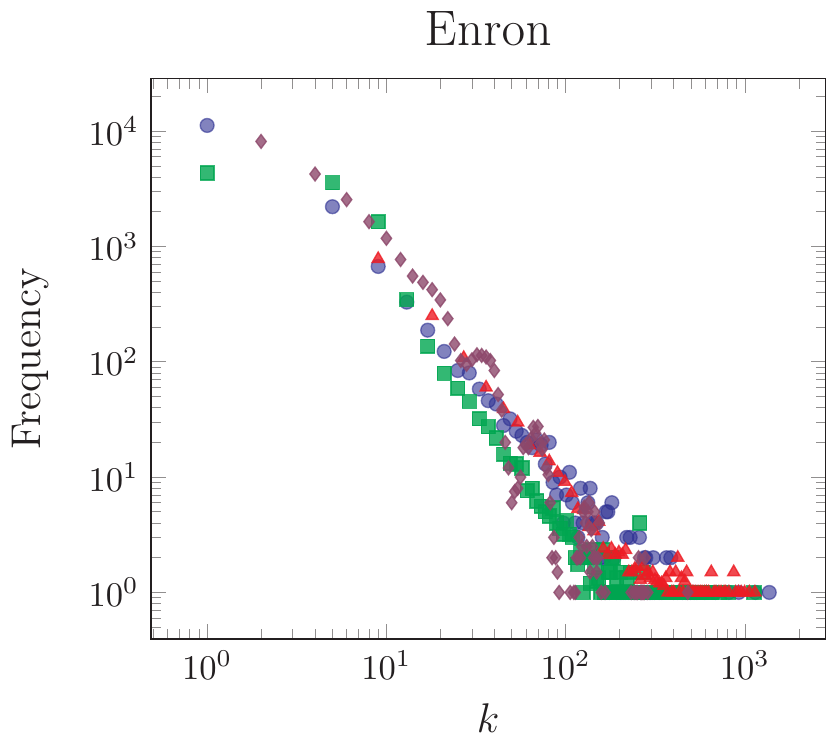}
    }
\end{minipage}    
\begin{minipage}[b]{0.23\textwidth}
    \resizebox{\textwidth}{!}{
        \includegraphics{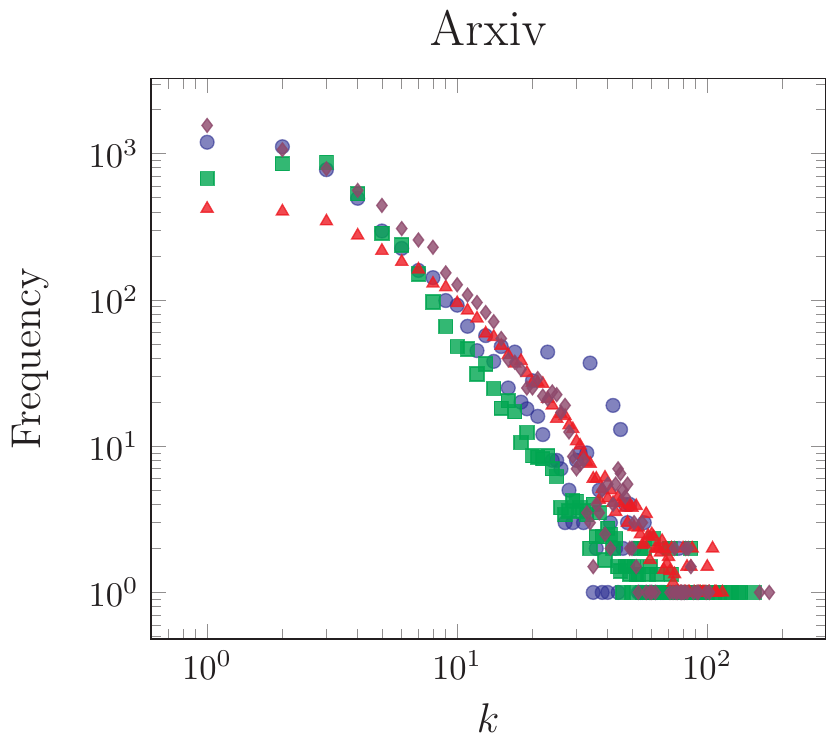}
    }
\end{minipage}%
\begin{minipage}[b]{0.23\textwidth}
    \resizebox{\textwidth}{!}{
        \includegraphics{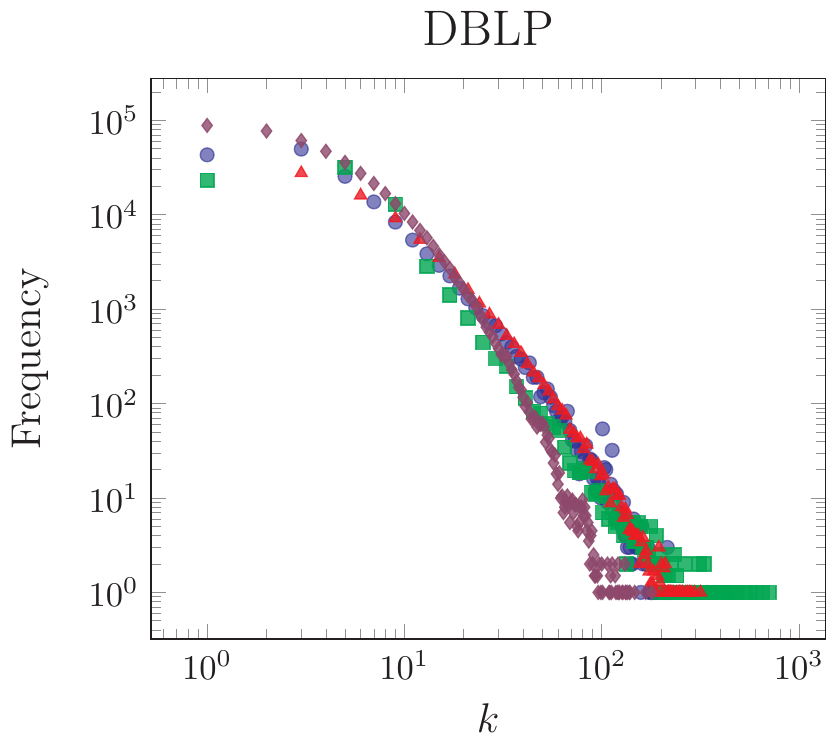}
    }
\end{minipage}
\resizebox{.37\linewidth}{!}{
    \includegraphics{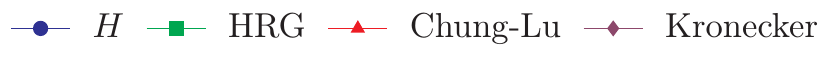}
}
\caption{Degree Distribution. Dataset graphs exhibit a power law degree distribution that is well captured by existing graph generators as well as HRG.}
\label{fig:real_degree}
\end{figure}

\begin{figure}[t]
\centering
\begin{minipage}[b]{0.23\textwidth}
    \resizebox{\textwidth}{!}{
        \includegraphics{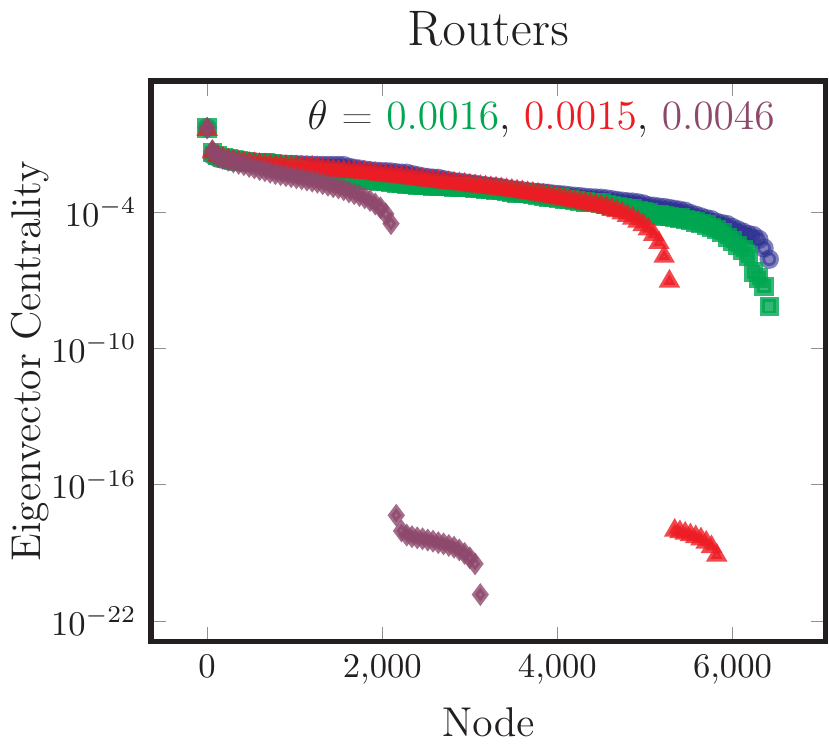}
    }
\end{minipage}%
\begin{minipage}[b]{0.23\textwidth}
    \resizebox{\textwidth}{!}{
        \includegraphics{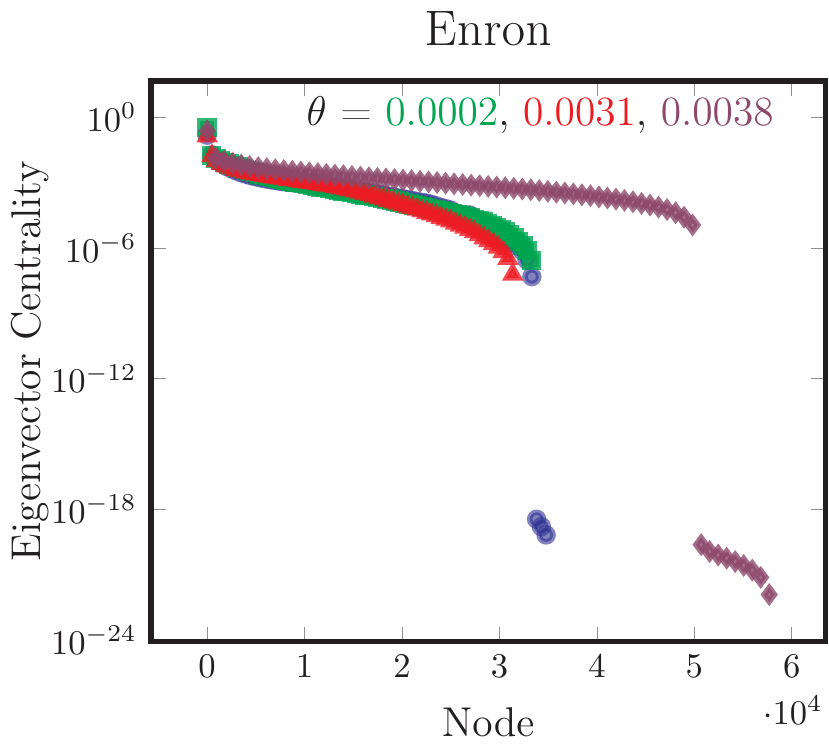}
    }
\end{minipage}    
\begin{minipage}[b]{0.23\textwidth}
    \resizebox{\textwidth}{!}{
        \includegraphics{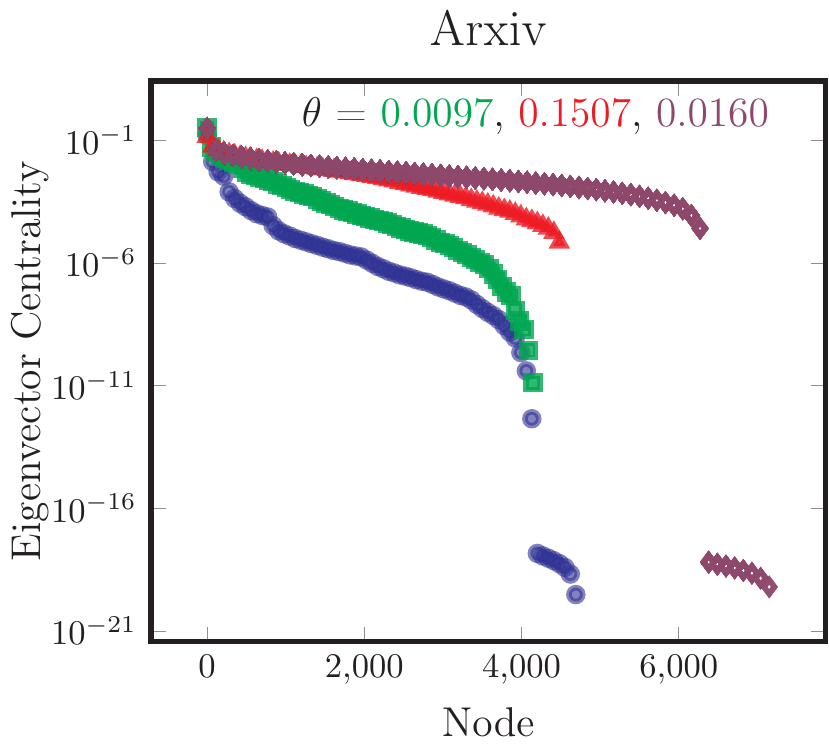}
    }
\end{minipage}%
\begin{minipage}[b]{0.23\textwidth}
    \resizebox{\textwidth}{!}{
        \includegraphics{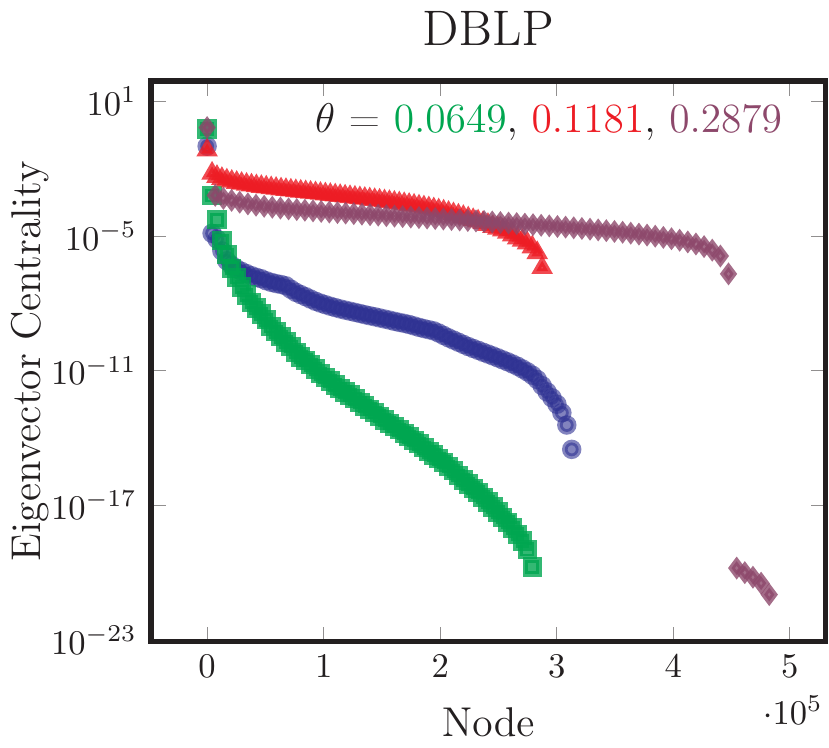}
    }
\end{minipage}
\resizebox{.37\linewidth}{!}{
    \includegraphics{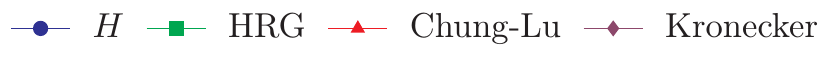}
}
\caption{Eigenvector Centrality. Nodes are ordered by their eigenvector-values along the x-axis. Cosine distance between the original graph and HRG, Chung-Lu and Kronecker models are shown at the top of each plot where lower is better. In terms of cosine distance, the eigenvector of HRG is consistently closest to that of the original graph.}
\label{fig:real_eig}
\end{figure}

\begin{figure}[t!]
\centering
\begin{minipage}[b]{0.23\textwidth}
    \resizebox{\textwidth}{!}{
        \includegraphics{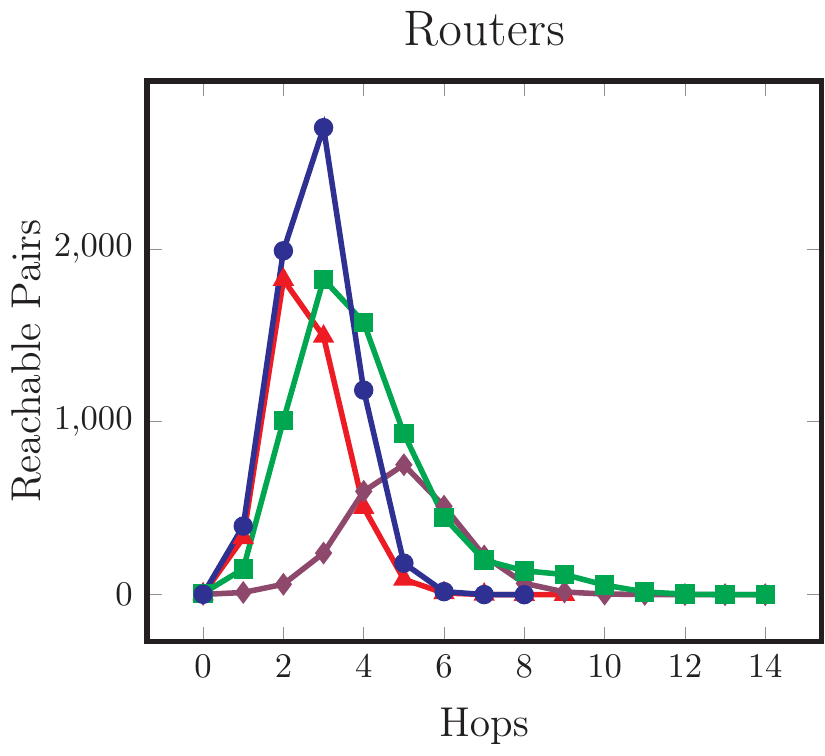}
    }
\end{minipage}%
\begin{minipage}[b]{0.222\textwidth}
    \resizebox{\textwidth}{!}{
        \includegraphics{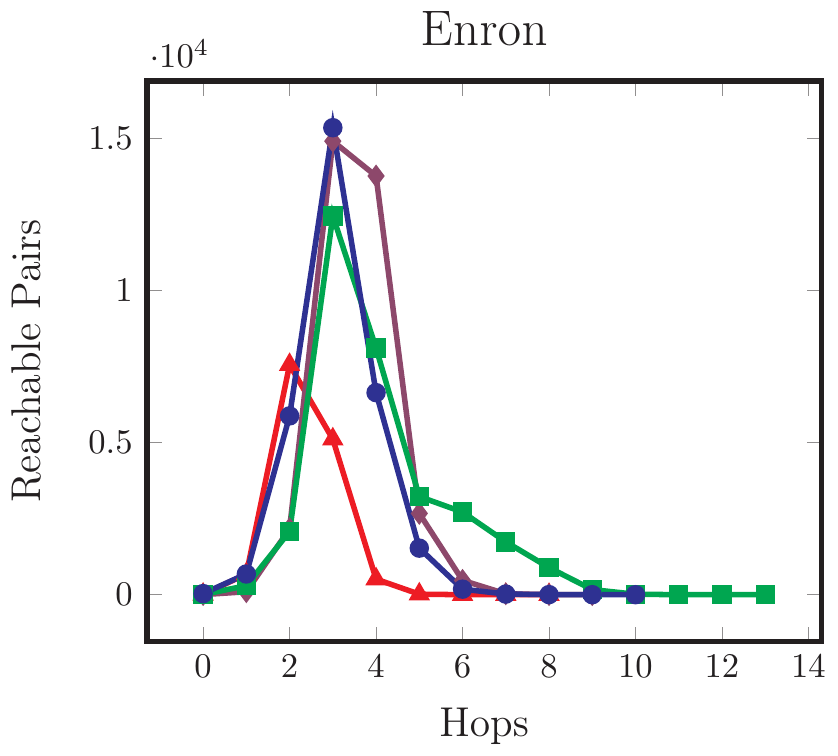}
    }
\end{minipage}    
\begin{minipage}[b]{0.23\textwidth}
    \resizebox{\textwidth}{!}{
        \includegraphics{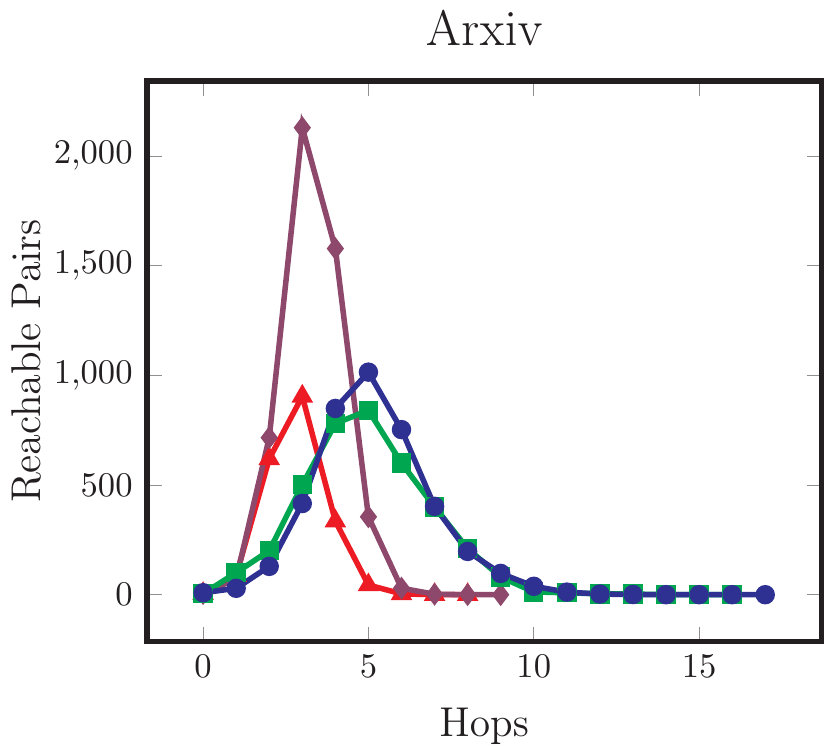}
    }
\end{minipage}%
\begin{minipage}[b]{0.23\textwidth}
    \resizebox{\textwidth}{!}{
        \includegraphics{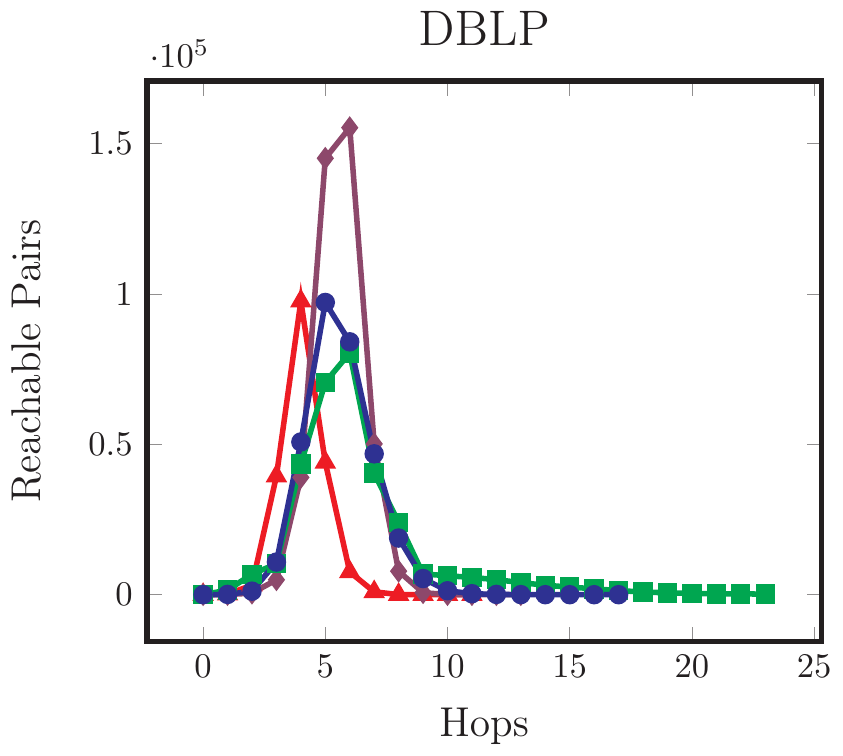}
    }
\end{minipage}
\resizebox{.37\linewidth}{!}{
    \includegraphics{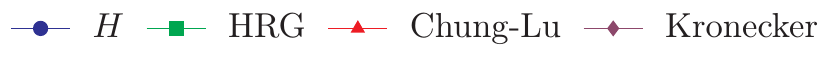}
}
\caption{Hop Plot. Number of vertex pairs that are reachable within $x$-hops. HRG closely and consistently resembles the hop plot curves of the original graph. }
\label{fig:real_hopplot}
\end{figure}

\myparagraph{Degree Distribution.} The degree distribution of a graph is the distribution of the number of edges connecting to a particular vertex. Barab\'{a}si and Albert initially discovered that the degree distribution of many real world graphs follows a power law distribution such that the number of nodes $N_d\propto d^{-\gamma}$ where $\gamma>0$ and $\gamma$ is typically between 2 and 3~\cite{barabasi1999emergence}.

Figure~\ref{fig:real_degree} shows the results of the degree distribution property on the four real world graphs ($frequency$ or $count$ as a function of degree $k$). Recall that the graph results plotted here and throughout the results section are the mean averages of 20 generated graphs. Each of the generated graphs is slightly different from the original graphs in their own way. As expected, we find that the power law degree distribution is captured by existing graph generators as well as the HRG model.

\myparagraph{Eigenvector Centrality.} The principal eigenvector is often associated with the centrality or ``value'' of each vertex in the network, where high values indicate an important or central vertex and lower values indicate the opposite. A skewed distribution points to a relatively few ``celebrity'' vertices and many common nodes.

The principal eigenvector value for each vertex is also closely associated with the PageRank and degree value for each node. Figure~\ref{fig:real_eig} shows the eigenvector scores for each node ranked highest to lowest in each of the four real world graphs. Because the x-axis represents individual nodes, Fig.~\ref{fig:real_eig} also shows the size difference among the generated graphs. HRG performs consistently well across all four types of graphs, but the log scaling on the y-axis makes this plot difficult to discern. To more concretely compare the eigenvectors, the pairwise cosine distance between eigenvector centrality of $H$ and the mean eigenvector centrality of each model's generated graphs appear at the top of each plot in order. HRG consistently has the lowest cosine distance followed by Chung-Lu and Kronecker.

\myparagraph{Hop Plot.} The hop plot of a graph shows the number of vertex-pairs that are reachable within $x$ hops. The hop plot, therefore, is another way to view how quickly a vertex's neighborhood grows as the number of hops increases. As in related work~\cite{leskovec2005graphs} we generate a hop plot by picking 50 random nodes and perform a complete breadth first traversal over each graph.

Figure~\ref{fig:real_hopplot} demonstrates that HRG graphs produce hop plots that are remarkably similar to the original graph. Chung-Lu performs rather well in most cases; Kronecker has poor performance on Arxiv and DBLP graphs, but still shows the correct hop plot shape.

\subsection{Canonical Graph Comparison}

The previous network properties primarily focus on statistics of the global network. However, there is mounting evidence which argues that the graphlet comparisons are the most complete way measure the similarity between two graphs~\cite{przulj2007biological,Ugander2013}. The graphlet distribution succinctly describes the number of small, local substructures that compose the overall graph and therefore more completely represents the details of what a graph ``looks like.'' Furthermore, it is possible for two very dissimilar graphs to have the same degree distributions, hop plots, etc., but it is difficult for two dissimilar graphs to fool a comparison with the graphlet distribution. 

\myparagraph{Graphlet Correlation Distance} Recent work from systems biology has identified a new metric called the Graphlet Correlation Distance (GCD). The GCD computes the distance between two graphlet correlation matrices -- one matrix for each graph~\cite{yaverouglu2015proper}. It measures the frequency of the various graphlets present in each graph, \ie, the number of edges, wedges, triangles, squares, 4-cliques, etc., and compares the graphlet frequencies between two graphs. Because the GCD is a distance metric, lower values are better. The GCD can range from $[0,+\infty]$, where the GCD is 0 if the two graphs are isomorphic.

\begin{figure}[t]
\centering
    \resizebox{.64\textwidth}{!}{
        \includegraphics{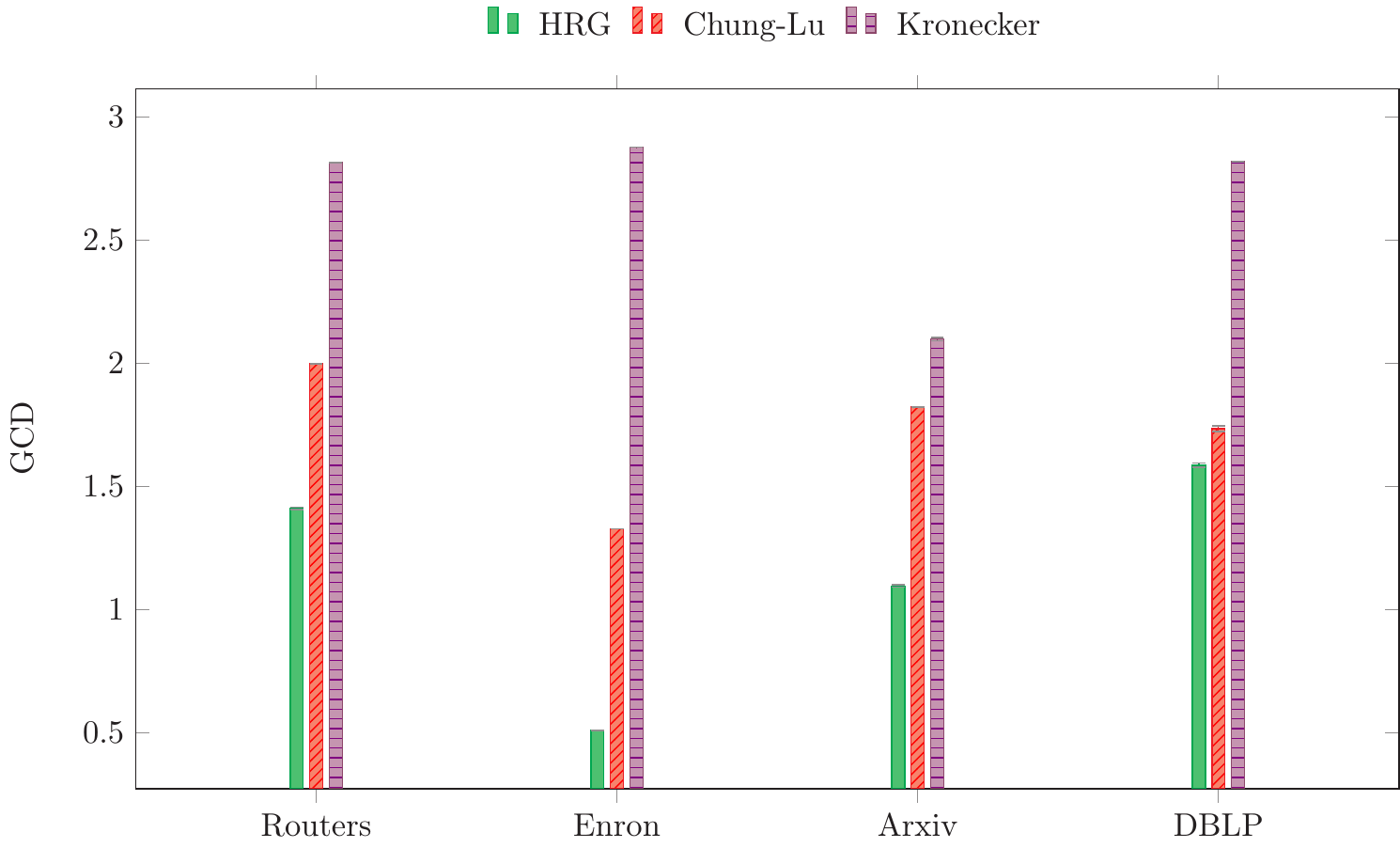}
    }
\caption{Graphlet Correlation Distance. A measure of the distance between the graphlet counts of both graphs, but also represents a canonical measure of graph similarity. Lower is better. }
\label{fig:gcd_real}
\end{figure}

We computed the GCD between the original graph and each generated graph. Figure~\ref{fig:gcd_real} shows the GCD results. Although they are difficult to see due to their small size, Fig.~\ref{fig:gcd_real} includes error bars for the 95\% confidence interval. The results here are clear: HRG significantly outperforms the Chung-Lu and Kronecker models.

The GCD opens a whole new line of network comparison methods that stress the graph generators in various ways. 

\subsection{Graph Extrapolation}

Recall that HRG learns the grammar from $k=4$ subgraph-samples from the original graph. In essence, HRG is extrapolating the learned subgraphs into a full size graph. This raises the question: if we only had access to a small subset of some larger network, could we use our models to infer a larger (or smaller) network with the same local and global properties? For example, given the 34-node Karate Club graph, could we infer what a hypothetical Karate Franchise might look like?

Using two smaller graphs, Zachary's Karate Club (34 nodes, 78 edges) and the protein-protein interaction network of \textit{S.~cerevisiae} yeast (1,870 nodes, 2,240 edges), we learned an HRG model with $k=1$ and $s=n$, \ie, no sampling, and generated networks of size-$n^*$ = 2x, 3x, \ldots, 32x. For the protein graph we also sampled down to $n^*=x/8$. Powers of 2 were used because the standard Kronecker model can only generate graphs of that size. The Chung-Lu model requires a size-$n^*$ degree distribution as input. To create the proper degree distribution we fitted a Poisson distribution ($\lambda = 2.43$) and a Geometric Distribution ($p = 0.29$) to Karate and Protein graphs respectively and drew $n^*$ degree-samples from their respective distributions. In all cases, we generated 20 graphs at each size-point.

\begin{figure}[t]
\centering
\begin{minipage}[b]{0.23\textwidth}
    \resizebox{\textwidth}{!}{
        \includegraphics{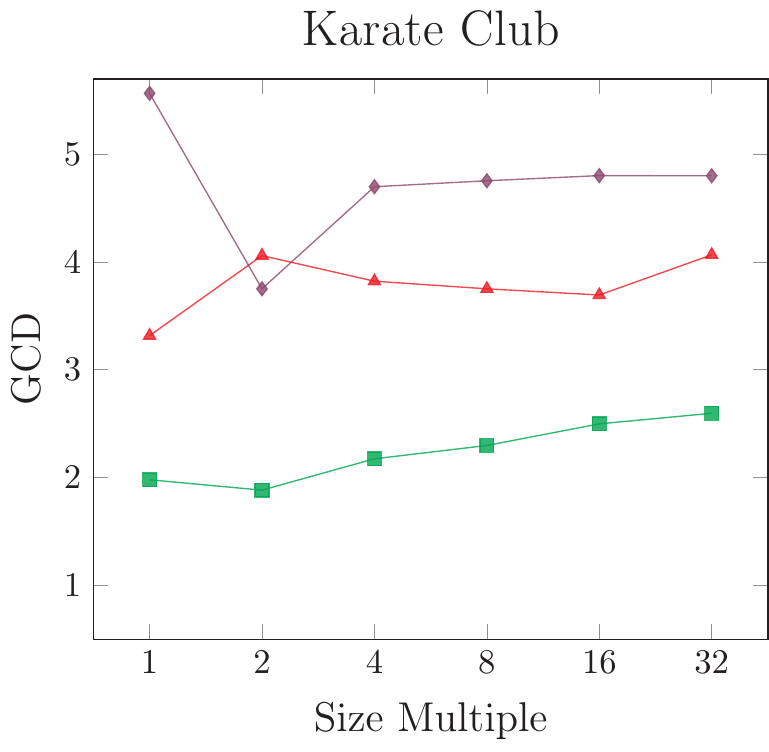}
    }
\end{minipage}
\begin{minipage}[b]{0.228\textwidth}
    \resizebox{\textwidth}{!}{
        \includegraphics{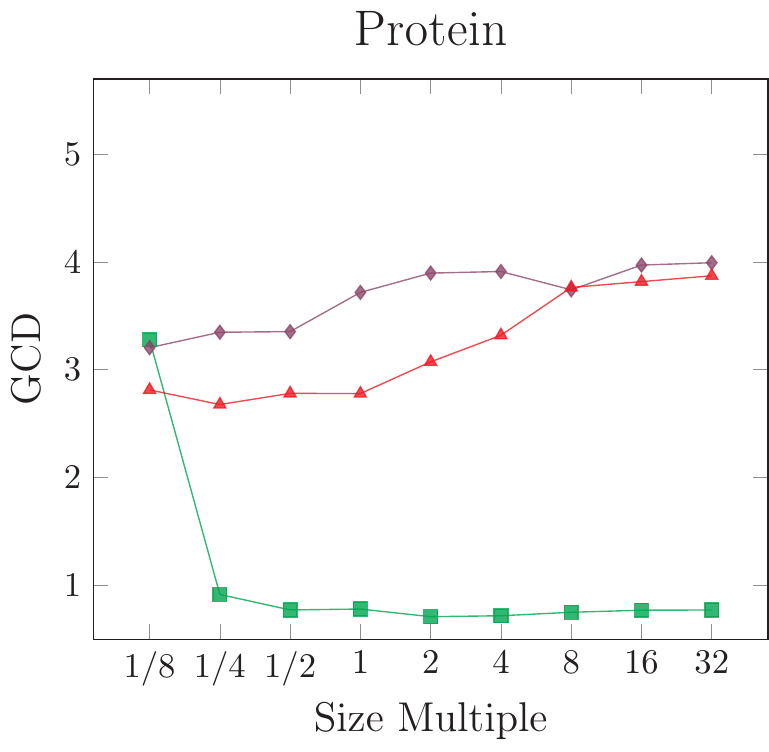}
    }
\end{minipage}
\\
\resizebox{.35\linewidth}{!}{
    \includegraphics{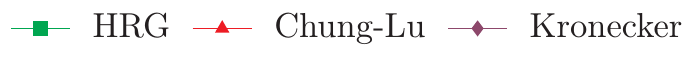}
}
\caption{GCD of graphs extrapolated in multiples up to 32x from two small graphs. HRG outperforms Chung-Lu and Kronecker models when generating larger graphs. Lower is better. }
\label{fig:xtrapol}
\end{figure}

Rather than comparing raw numbers of graphlets, the GCD metric compares the \textit{correlation} of the resulting graphlet distributions. As a result, GCD is largely immune to changes in graph size. Thus, GCD is a good metric for this extrapolation task. Figure~\ref{fig:xtrapol} shows the mean GCD score and 95\% confidence intervals for each graph model. Not only does HRG generate good results at $n^*=1$x, the GCD scores remain mostly level as $n^*$ grows.

\subsection{Sampling and Grammar Complexity}

We have shown that HRG can generate graphs that match the original graph from $k=4$ samples of $s=500$-node subgraphs. If we adjust the size of the subgraph, then the size of the clique tree will change causing the grammar to change in size and complexity. A large clique tree ought to create more rules and a more complex grammar, resulting in a larger model size and better performance; while a small clique tree ought to create fewer rules and a less complex grammar, resulting in a smaller model size and a lower performance.

\begin{figure}[t]
\centering
    \resizebox{.64\textwidth}{!}{
        \includegraphics{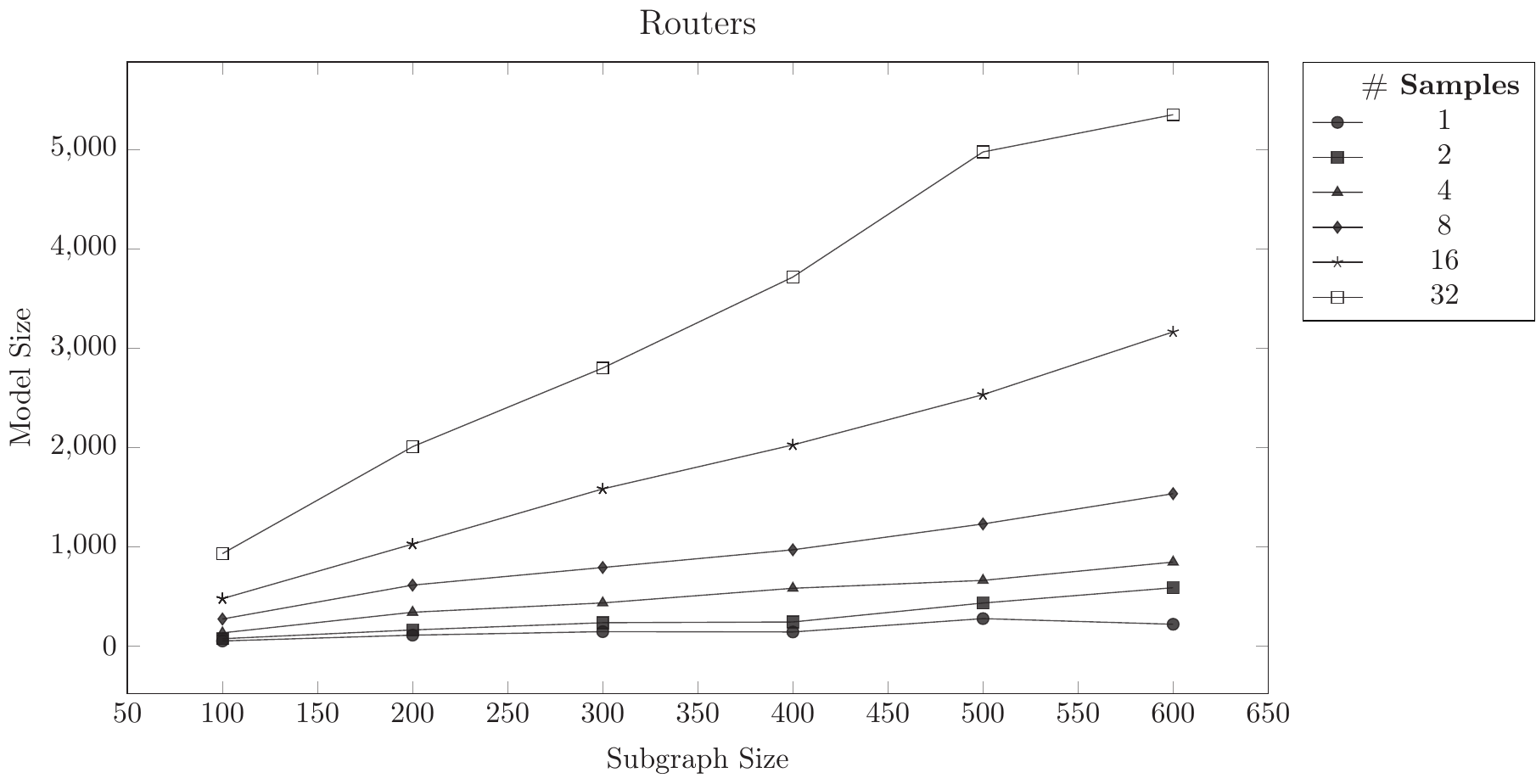}
    }

\caption{HRG model size as the subgraph size $s$ and the number of subgraph samples $k$ varies. The model size grows linearly with $k$ and $s$.}
\label{fig:grammarsize}
\end{figure}

To test this hypothesis we generated graphs by varying the number of subgraph samples $k$ from 1 to 32, while also varying the size of the sampled subgraph $s$ from 100 to 600 nodes. Again, we generated 20 graphs for each parameter setting. Figure~\ref{fig:grammarsize} shows how the model size grows as the sampling procedure changes on the Internet Routers graph. Plots for other graphs show a similar growth rate and shape, but are omitted due to space constraints. 

To test the statistical correlation we calculated Pearson's correlation coefficient between the model size and sampling parameters. We find that the $k$ is slightly correlated with the model size on Routers ($r=0.31$, $p=0.07$), Enron ($r=0.27, p=0.09$), ArXiv ($r=0.21, p=0.11$), and DBLP ($r=0.29$, $p=0.09$). Furthermore, the choice of $s$ affects the size of the clique tree from which the grammars are inferred. So its not surprising that $s$ is highly correlated with the model size on Routers ($r=0.64), Enron ($r=0.71), ArXiv ($r=0.68$), and DBLP ($r=0.54$) all with $p\ll 0.001$.

Because we merge identical rules when possible, we suspect that the overall growth of the HRG model follows Heaps law~\cite{heaps1978information}, \ie, that the model size of a graph can be predicted from its rules; although we save a more thorough examination of the grammar rules as a matter for future work.

\subsubsection{Model size and Performance}

One of the disadvantages of the HRG model, as indicated in Fig.~\ref{fig:grammarsize}, is that the model size can grow to be very large. But this again begs the question: do larger and more complex HRG models result in improved performance?

\begin{figure}[t]
\centering
\begin{minipage}[b]{0.23\textwidth}
    \resizebox{\textwidth}{!}{
        \includegraphics{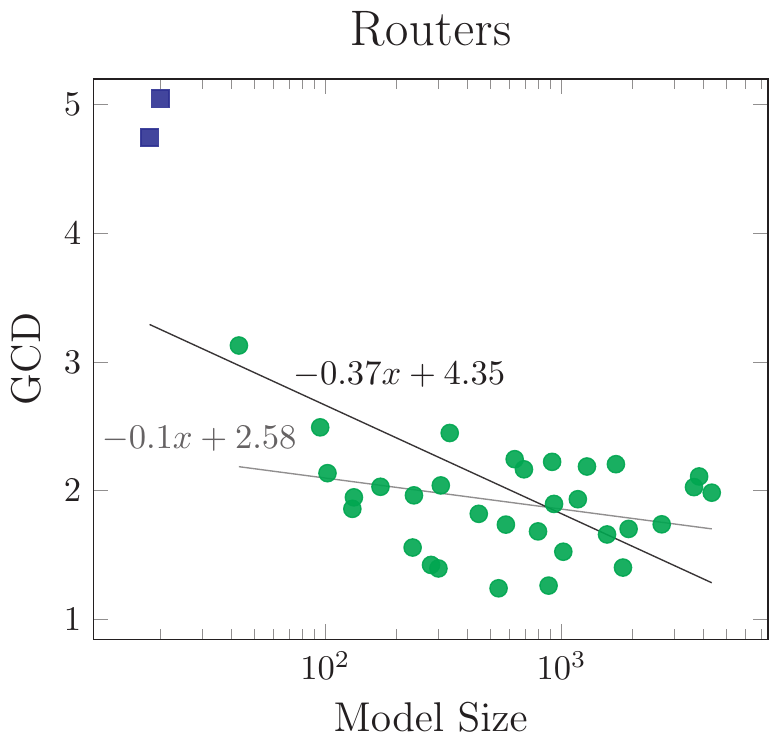}
    }
\end{minipage}
\begin{minipage}[b]{0.23\textwidth}
    \resizebox{\textwidth}{!}{
        \includegraphics{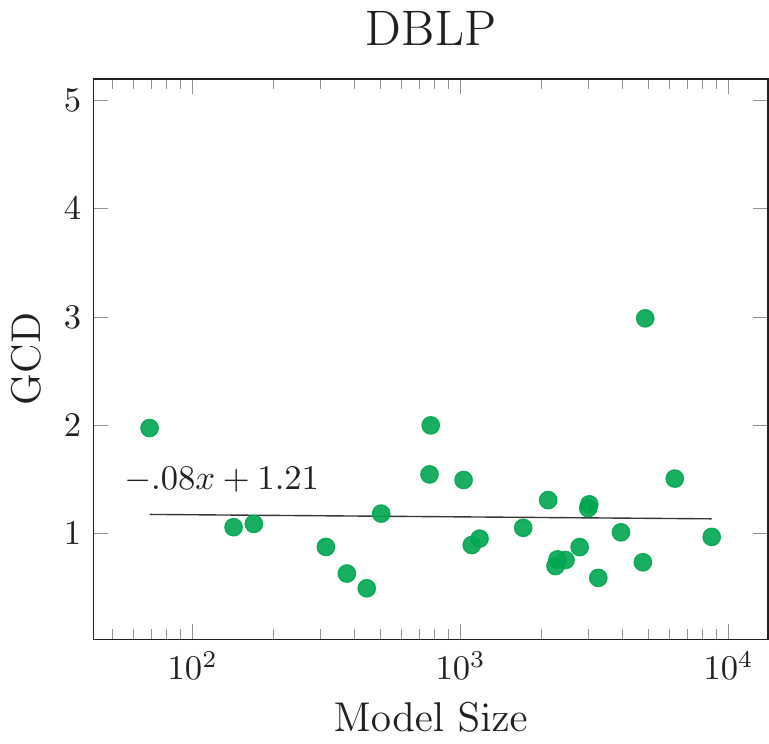}
    }
\end{minipage}
\caption{GCD as a function of model size. We find a slightly negative relationship between model size and performance, but with quickly diminishing returns. Best fit lines and their equations are shown; the shorter fit line in the Routers plot ignores the square outlier points. }
\label{fig:size_score}
\end{figure}

To answer this question we computed the GCD distance between the original graph and graphs generated by varying $k$ and $s$. Figure~\ref{fig:size_score} illustrates the relationship between model size and the GCD. We use the Router and DBLP graphs to shows the largest and smallest of our dataset; other graphs show similar results, but their plots are omitted due to of space. Surprisingly, we find that the performance of models with only 100 rules is similar to the performance of the largest models. In the Router results, two very small models with poor performance had only 18 and 20 rules each. Best fit lines are drawn to illustrate the axes relationship where negative slope indicates that larger models generally perform better. Outliers can dramatically affect the outcome of best fit lines, so the faint line in the Routers graph shows the best fit line if we remove the two square outlier points. Without removing outliers, we find only a slightly negative slope on the best fit line indicating only a slight performance improvement between HRG models with 100 rules and HRG models with 1,000 rules. Pearson's correlation coefficient comparing GCD and model size similarly show slightly negative correlations on Routers ($r=-0.12$, $p=0.49$), Enron ($r=-0.09, p=0.21$), ArXiv ($r=0.04, p=0.54$), and DBLP ($r=-0.08$, $p=0.62$)

\subsubsection{Runtime Analysis Revisited \label{sss:runtime}}

The overall execution time of the HRG model is best viewed in two parts: (1) rule extraction, and (2) graph generation.

We previously identified the runtime complexity of the rule extraction process to be $O(m\cdot\Delta)$. However, this did not include $k$ samples of size-$s$ subgraphs. So, when sampling with $k$ and $s$, we amend the runtime complexity to be $O(k\cdot m \cdot \Delta)$ where $m$ is bounded by the number of hyperedges in the size-$s$ subgraph sample and $\Delta\le s$. Graph generation requires a straightforward application of rules and is linear in the number of edges in the output graph.

\begin{figure}[t]
\centering
\begin{minipage}[b]{0.23\textwidth}
    \resizebox{\textwidth}{!}{
        \includegraphics{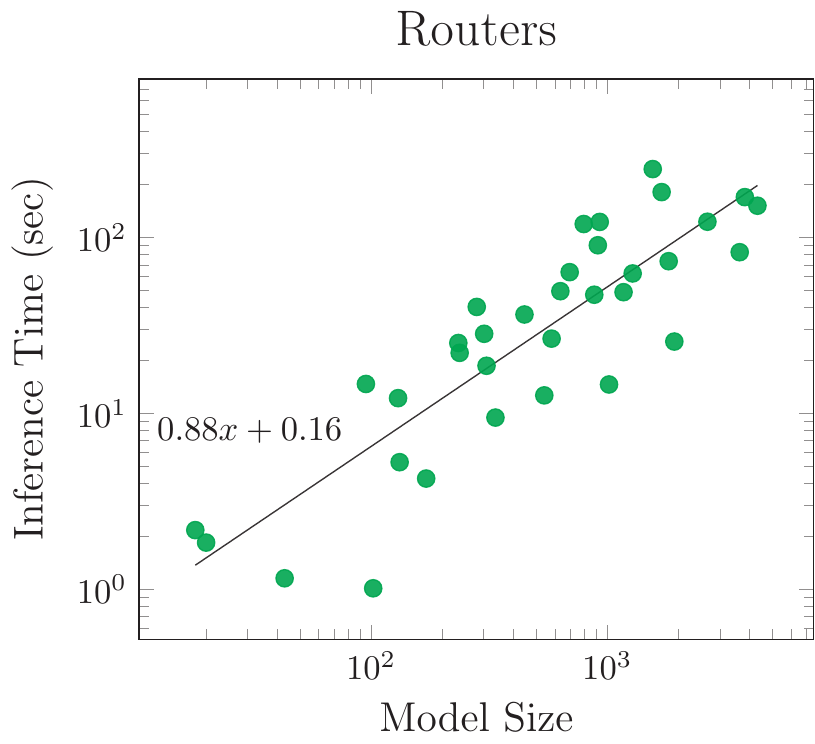}
    }
\end{minipage}
\begin{minipage}[b]{0.23\textwidth}
    \resizebox{\textwidth}{!}{
        \includegraphics{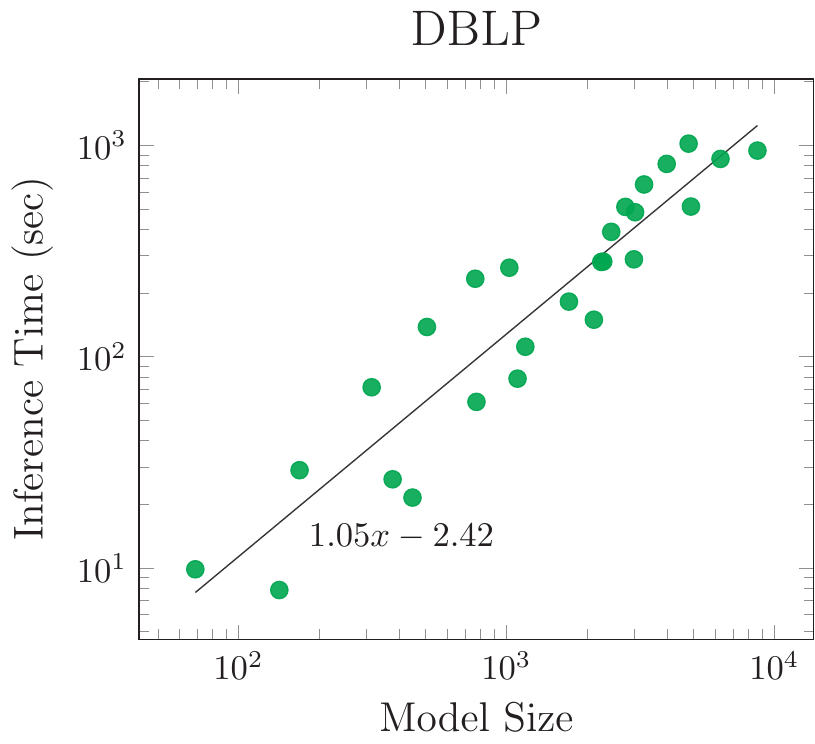}
    }
\end{minipage}
\caption{Total inference runtime (\ie, clique tree creation and rule extraction) as a function of model size. Best fit lines on the log-log plot show that the execution time grows linearly with the model size. }
\label{fig:size_time}
\end{figure}

All experiments were performed on a modern consumer-grade laptop in an unoptimized, unthreaded python implementation. We recorded the extraction time while generating graphs for the size-to-GCD comparison in the previous section. Although the runtime analysis gives theoretical upper bounds to the rule extraction process, Fig.~\ref{fig:size_time} shows that the extraction runtime is highly correlated to the size of the model in Routers ($r=0.68$), ArXiv ($r=0.91$), Enron ($r=0.88$), and DBLP ($r=0.94$) all with $p\ll 0.001$. Simply put, more rules require more time, but there are diminishing returns. So it may not be necessary to learn complex models when smaller HRG models tend to perform reasonably well.

\subsection{Graph Generation Infinity Mirror}

Lastly, we characterize the robustness of graph generators by introducing a new kind of test we call the \textit{infinity mirror}~\cite{aguinaga16infinity}. One of the motivating questions behind this idea was to see if HRG holds sufficient information to be used as a reference itself. In this test, we repeatedly learn a model from a graph generated by the an earlier version of the same model. For HRG, this means that we learn a set of production rules from the original graph $H$ and generate a new graph $H^*$; then we set $H\gets H^*$ and repeat thereby learning a new model from the generated graph recursively. We repeat this process ten times, and compare the output of the tenth recurrence with the original graph using GCD. 

We expect to see that all models degenerate over 10 recurrences because graph generators, like all machine learning models, are lossy compressors of information. The question is, how quickly do the models degenerate and how bad do the graphs become?

\begin{figure}[t]
\centering
\begin{minipage}[b]{0.235\textwidth}
    \resizebox{\textwidth}{!}{
        \includegraphics{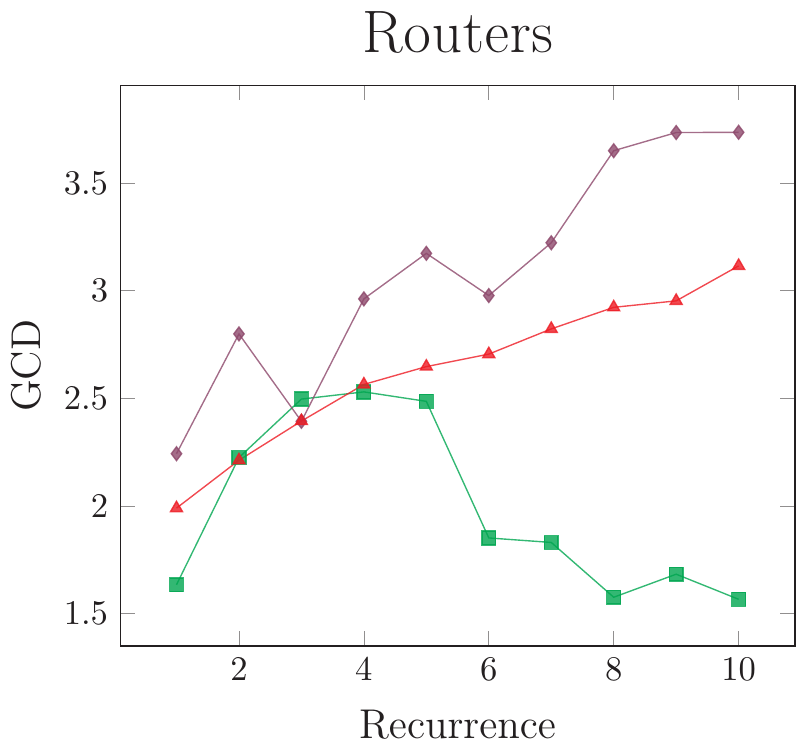}
    }
\end{minipage}
\begin{minipage}[b]{0.235\textwidth}
    \resizebox{\textwidth}{!}{
        \includegraphics{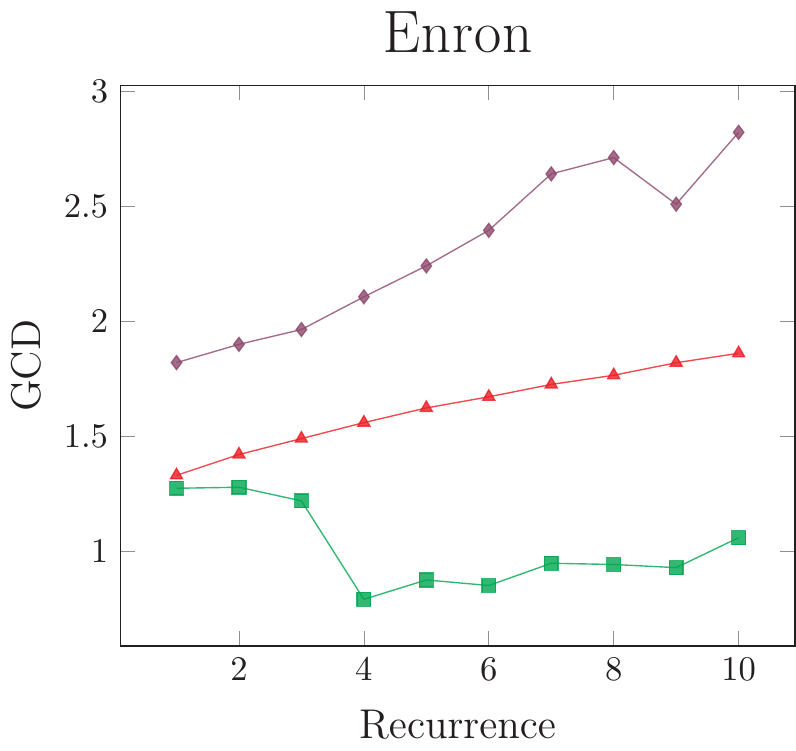}
    }
\end{minipage}
\begin{minipage}[b]{0.235\textwidth}
    \resizebox{\textwidth}{!}{
        \includegraphics{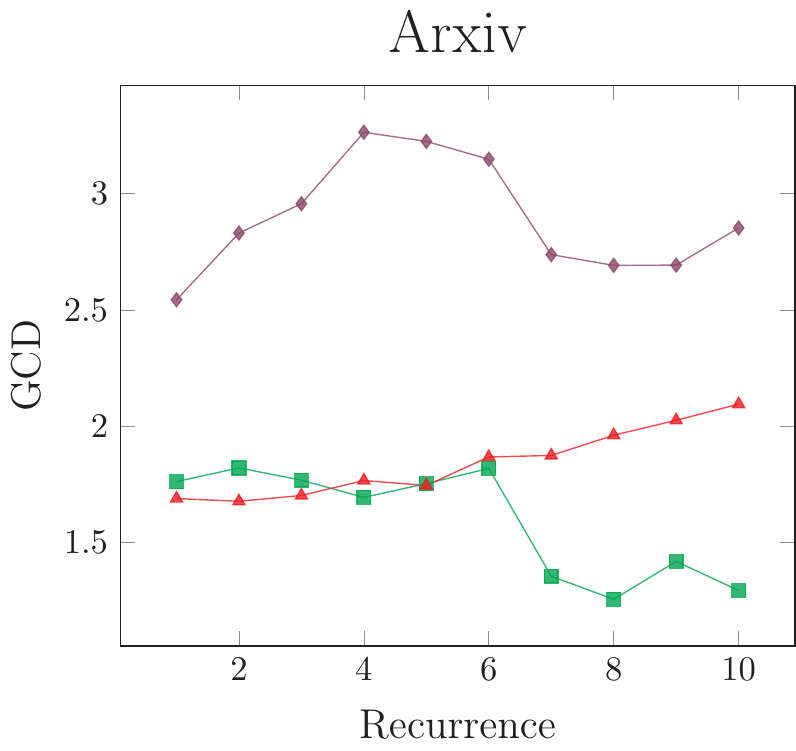}
    }
\end{minipage}
\begin{minipage}[b]{0.235\textwidth}
    \resizebox{\textwidth}{!}{
        \includegraphics{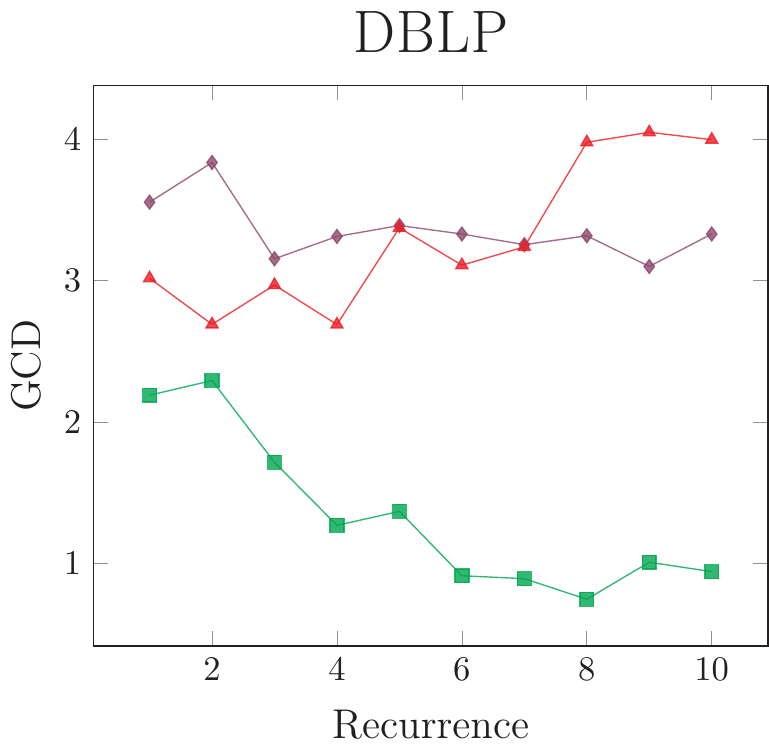}
    }
\end{minipage}
\resizebox{.37\linewidth}{!}{
    \includegraphics{./arxiv-figure27}
}
\caption{Infinity Mirror. Unlike Kronecker and Chung-Lu models, HRG does not degenerate as its model is applied repeatedly. Lower is better.}
\label{fig:inf_mir_gcd}
\end{figure}

Figure~\ref{fig:inf_mir_gcd} shows the GCD scores for the HRG, Chung-Lu and Kronecker models at each recurrence. Surprisingly, we find that HRG stays steady, and even improves its performance while the Kronecker and Chung-Lu models steadily decrease their performance as expected. We do not yet know why HRG improves performance in some cases. Because GCD measures the graphlet correlations between two graphs, the improvement in GCD may be because HRG is implicitly honing in on rules that generate the necessary graph patterns. Yet again, further work is needed to study this important phenomenon.

\section{Conclusions and Future Work}

In this paper we have shown how to use clique trees (also known as junction trees, tree decomposition, intersection trees) constructed from a simple, general graph to learn a hyperedge replacement grammar (HRG) for the original graph. We have shown that the extracted HRG can be used to reconstruct a new graph that is isomorphic to the original graph if the clique tree is traversed during reconstruction. More practically, we show that a stochastic application of the grammar rules creates new graphs that have very similar properties to the original graph. The results of graphlet correlation distance experiments, extrapolation and the infinity mirror are particularly exciting because our results show a stark improvement in performance over existing graph generators.

In the future, we plan to investigate differences between the grammars extracted from different types of graphs; we are also interested in exploring the implications of finding two graphs which have a large overlap in their extracted grammars. Among the many areas for future work that this study opens, we are particularly interested in learning a grammar from the actual growth of some dynamic or evolving graph. Within the computational theory community there has been a renewed interest in quickly finding clique trees of large real world graphs that are closer to optimal. Because of the close relationship of HRG and clique trees shown in this paper, any advancement in clique tree algorithms could directly improve the speed and accuracy of graph generation.

Perhaps the most important finding that comes from this work is the ability to interrogate the generation of substructures and subgraphs within the grammar rules that combine to create a holistic graph. Forward applications of the technology described in this work may allow us to identify novel patterns analogous to the previously discovered triadic closure and bridge patterns found in real world social networks. Thus, an investigation in to the nature of the extracted rules and their meaning (if any) is a top priority.

We encourage the community to explore further work bringing HRGs to attributed graphs, heterogeneous graphs and developing practical applications of the extracted rules. Given the current limitation related to the growth in the number of extracted rules as well as the encouraging results from small models, we are also looking for sparsification techniques that might limit the model's size while still maintaining performance. 

\section{Acknowledgements}
This work is supported by the Templeton Foundation under grant FP053369-M/O.

\end{document}